\documentclass[12pt]{article}
\usepackage{amsmath}
\usepackage{graphicx}
\usepackage{enumerate}
\usepackage{natbib}
\usepackage{url}
\usepackage{color}
\usepackage{float}
\usepackage[utf8]{inputenc}
\usepackage{amsthm} 
\usepackage{amssymb}
\usepackage{multirow}
\usepackage{latexsym}
\usepackage{enumitem}

\newcommand{\E}{{\rm E}}

\def\d{{\mathrm d}}

\newtheorem{remark}{Remark}

\newcommand{\blind}{1}

\usepackage{float}
\def\d{{\mathrm d}} 

\begin{document}

\def\spacingset#1{\renewcommand{\baselinestretch}%
	{#1}\small\normalsize} \spacingset{1}


\if1\blind
{
	\title{\bf Low quality exposure and point processes with a view to the first phase of a pandemic 
	}
	\author{
		Mar{\'\i}a Luz G\'amiz\thanks{Corresponding author: mgamiz@ugr.es} \\ Department of Statistics and O.R., University of Granada, Spain\\ \\
		Enno Mammen\\ Institute of Applied Mathematics, Heidelberg University, Germany\\ \\
		Mar{\'\i}a Dolores Mart{\'\i}nez-Miranda
		\\ Department of Statistics and O.R., University of Granada, Spain\\ \\
		Jens Perch Nielsen \\
		Bayes Business School, City, University of London, London, UK
	}
	\maketitle
} \fi

\if0\blind
{
	\bigskip
	\bigskip
	\bigskip
	\begin{center}
		{\LARGE\bf  Low quality exposure and point processes with a view to the first phase of a pandemic}
	\end{center}
	\medskip
} \fi

\bigskip
\abstract{In the early days of  a pandemic there is no time for complicated data collection. One needs a simple cross-country benchmark approach based on robust data that is easy to understand and easy to collect. The recent pandemic has shown us what early available pandemic data might look like, because statistical data was published every day in standard news outlets in many countries.
This paper provides new methodology for the analysis of data where exposure is only vaguely understood and where the very definition of exposure might change over time. The exposure of poor quality is used to analyse and forecast events. Our example of such exposure is daily infections during a pandemic and the events are number of new infected patients in hospitals every day.
Examples are given with French Covid-19 data on hospitalized patients and numbers  of infected.
}

\bigskip


\maketitle

\newpage
\spacingset{1.5} 

\section{Introduction}
\label{sec:intro}

During a global pandemic, development is likely to be chaotic, requiring the collection of information and knowledge from various environments daily. The outbreak might shift from one country to another, and definitions and measurements change frequently. For instance, hospitalization criteria may evolve dynamically, while the daily number of hospitalized individuals can be reliably recorded in many countries. On the other hand, the definition of exposure, such as daily infections during a pandemic, might be vaguely understood and subject to change over time. Standardization and easy exchange of both input and output are crucial in the methodology used. Data collection time, communication and knowledge sharing across communities, quick data processing and easily understood output are important elements when helping society understand a developing pandemic.

 \cite{Richardson:22} argues that the new challenges posed by the pandemic have highlighted how the ubiquitous nature of statistical thinking gives us the capacity to understand and model complex contexts and dynamics, dealing with constraints of agility, responsiveness and societal responsibilities. When good quality data are available it is possible to work with complicated statistical models.   \cite{Giudici:et:al:23} proposes a model using a large dataset that includes spatio-temporal, mobility and socio-demographic covariates; \cite{Samyak:Palacios:23} {{applies}} evolutionary trees based on different subsamples of SARS-CoV-2 molecular sequences across different states of the US; \cite{Stokell:et:al:21} compares a collection of predictive models with a purpose of establishing spacial clusters in the US; in \cite{Jiang:et:al:22} the dynamics of the infection is accounted by a quantile-based approach which is robust to outliers and captures heterocedasticity; \cite{Millimet:Parmeter:22} {{extends} a model based on the SIR framework to allow for underreporting number of cases and deaths.

 During the course of the Covid-19 pandemic{{,} data analyses have shown to be an impressive tool for decision makers. But at the first phase of a pandemic, decisions have to be made rapidly. At that stage, there is not enough time to gather detailed data. This  paper provides a new methodology introducing a dynamic approach to monitor a pandemic with available data. The data used reflect the type of data almost every news{{-}}reading person learned to know during the Covid-19 pandemic. 
 The simple structure of this data is a mathematical statistical challenge. The missing link between origin and end events in the available data calls for new strategies. The new methodology of this paper is introducing a dynamic extension in various directions of the recent paper \cite{Gamiz:etal:22} that provides a new technique solving this new missing-link data problem for survival analysis, and uses it on Covid-19 pandemic data. 

There is a strong connection between our work and the planning presented in \cite{Garrett:23} on behalf of the Royal Statistical Society, designed to prepare for future pandemics. The approach proposed in this paper can serve as a benchmark methodology at the early stages of a pandemic. In \cite{Garrett:23}, the Royal Statistical Society emphasizes the importance of having a robust dynamic approach capable of providing reliable early results even when detailed data are scarce. This is precisely the purpose of the present paper, which aims to predict measured infections and their dynamic impact on hospitalizations.

 Our approach is able to include expert knowledge input on the reproduction number, see \cite{Storvik:etal:22}, \cite{Pellis:et:al:22}, \cite{Park:et:al:20}, and others.  With our approach, experts on the {reproduction number} do not any longer need to be experts in the overall scientific modelling of the problem. Their contribution can be fed into the system as expert knowledge. Our approach can also be used to test the accuracy of past reproduction numbers from experts.
 
 In this paper, the focus is on the infection process and hospitalizations of infected patients.  When modelling the transition rate from number of infected individuals to number of hospitalized, the number of individuals involved are based on some subsample of the population and even with a dynamic criteria for selecting this subsample, i.e. the number of positive tested.  
 
 The practical applications of the methodology developed in this paper are not limited to monitoring  a pandemic. In society, forecasting events before they happen is common and exposure data is often of low quality and dynamically changing over time. There are many {examples} besides hospital admissions during a pandemic{,} including crimes, insurance fraud and tax avoidance. 
In police crime cases, the quality of police actions in collecting local information on potential suspicious events or individuals may vary.  Also for insurance fraud or tax avoidance, the quality of data collection can vary over time. A dynamic approach is therefore necessary to adjust for the time dependency in the low quality exposure information.

Let $N_2$ be a counting process measuring the number of events over time and let $N_1$ be a vaguely connected stochastic process that might impact future values of $N_2$. {Here} $N_1$ does not need to be a counting process. One key assumption in our approach is that there is a clear correlation of values of $N_1$ in neighbouring time points: if there is a lot going on at one time point, this will lead to high activity in the near future as well. We will model the effect from one time point of $N_1$ to another via a dynamic time stochastic process, equal to or related to a {non-stationary} Hawkes process.  
At every time point $t$, this dynamic process is started and scaled with $N_1(t)$ to provide future contributions to $N_1$. The events $N_2$ are modelled via a modification of a Hawkes process starting at every time point $t$ and scaled with $N_1(t)$.  
In the pandemic case, $N_2$ is the number of entrances into hospital{,} or $N_2$ is simply equal to $N_1$, where $N_1$ is the number of registered infections.

The modelling of $N_1$ and $N_2$ provides a model where it is possible to forecast the future values of {$N_1$ and } $N_2$, given the information available in $N_1$. However, if the model should be general enough to {monitoring the development of the processes}, then it should be possible to incorporate information or prior knowledge from other sources than just $N_1$ and $N_2$. The approach of this paper is able to incorporate such extra knowledge in a very simple way: via a single number that provides an intervention on the forecasting of $N_1$.
That number represents the expected change of the future rate of events, which we denote by $C_{t^*,h}$, where $t^*$ is the most recent estimation time, $t^*+h$, with $h>0$,  is the forecasting horizon. If there would be no intervention{,} and no useful prior knowledge{,} then {$C_{t^*,h}$} is simply equal to one. If an intervention or other prior knowledge is provided{,} indicating lower future events of $N_2${,} then {$C_{t^*,h}$} is set to a number lower than one, and vice versa when intervention or prior knowledge {indicates} an increase in future events. In the pandemic case, we will show that {$C_{t^*,h}$} corresponds to the well{-}known reproduction number, that was a daily news information in many countries during the most severe parts of the pandemic. The advantage of the approach of this paper is that it provides the difficult algorithms dealing with the modelling of the number of entrances to hospital{,} based on the easy-to-collect and easy-to-understand information present in $N_1$ and $N_2$. Extra information to monitor and forecast can be incorporated via the well{-}known reproduction number. 

Our approach when modelling $N_1$ can be seen as a probabilistic queueing theory approach, see for example the related paper \cite{Goldenshluger:Koops:19}, or it can be understood as a missing data problem, because we do not know the exact link between one infected and another infected, a link we would normally use while estimating stochastic processes, see \cite{Gamiz:etal:22} for details. There is {a} missing data interpretation also when considering the prediction or forecasting of $N_2$ based on $N_1$. These missing data problems are new and do not match EM-algorithms or other well known missing data approaches, see for example  \cite{Heckman:79},  \cite{Rubin:96}, \cite{Zhao:etal:20},  \cite{Liu:Hu:22}, \cite{Little:Rubin:19}, \cite{Kim:Shao:21}, and \cite{Gamiz:etal:22}. 

The  main contributions of this paper are summarized in the following. First we describe the modelling of the vaguely defined exposure measure $N_1$ as a Hawkes process, or as a related process, and provide an iterative algorithm to estimate the entering quantities. We also provide theoretical justification for the estimators. Then we model the relationship between $N_2$ and $N_1$ via a modification of a Hawkes process, and provide another iterative algorithm for estimation, along with a theoretical justification. {We propose a bootstrap procedure to quantify the uncertainty of our estimators and forecasts.} To provide forecasts we introduce our principles of incorporating extra information or expert knowledge into our system via our simple number $C_{t^*,h}$. Finally, we provide a complete case study including the modelling of $N_1$ and $N_2$ and the use of extra information or expert knowledge. Our example is the Covid-19 pandemic in France and the expert knowledge is the weekly published reproduction number.

{
The organisation of the paper is as follows.  Section \ref{sec:model} provides methodology to monitor the feedback loop of the vaguely defined exposure process {$N_1$}, 
and  to connect the events of interest to the vaguely defined exposure process. 
Section \ref{sec:estim} introduces estimators for the two-dimensional transition intensities described in Section \ref{sec:model} in two different scenarii: when individual follow-up data are available; and, when infection and hospitalization timelines are not directly observable from the data.
Section \ref{sec:forecast} shows how the methodology of Section \ref{sec:model} can be used for forecasting, adding only one number from expert knowledge, or other sources. Section \ref{sec:application} goes through the methodology of Section \ref{sec:model}, using available French Covid-19 data. Section \ref{sec:fore_France} shows how to combine the methodology of Section \ref{sec:forecast} with the results of Section \ref{sec:application} to forecast hospitalizations in a developing pandemic. 
Section \ref{sec:comparison} presents a comparison with recent related works.
Section \ref{sec:conc} provides the conclusions of the paper and establishes connections with some of the recent methodological papers on the Covid-19 pandemic. Additional material is provided in the appendices consisting of the asymptotic analysis (Appendix \ref{sec:theo1}, further details on the derivation of the local linear estimators (Appendix \ref{ap:estim}), a bootstrap test for stationary delay distribution (Appendix \ref{app:test}), a practical correction for daily variations in the data (Appendix \ref{app:koyama}), and a brief simulation study (Appendix \ref{sec:simu}).  Finally,  additional data analyses are provided in the Supplementary Material, including the French Covid-19 case using weekly data, as well as a preliminary analysis of the UK case. 
}


\section{Model formulation}\label{sec:model}

In this section we first focus on the vaguely defined exposure process $N_1$, and then we turn to the modelling of $N_2$ that is the process we really want to model. When modelling and forecasting $N_2$, we will take advantage of the available information of $N_1$ and its modelling. 

 { Figure \ref{fig:transitions} shows  the feedback loop from infection to infection as well as  the transition from infected to hospitalization. We denote by $\lambda_1$ the infection rate, and $\lambda_2$ is the hospitalization rate. }

\begin{figure}[htp]
	\centering
	\makebox{\includegraphics[scale=0.35]{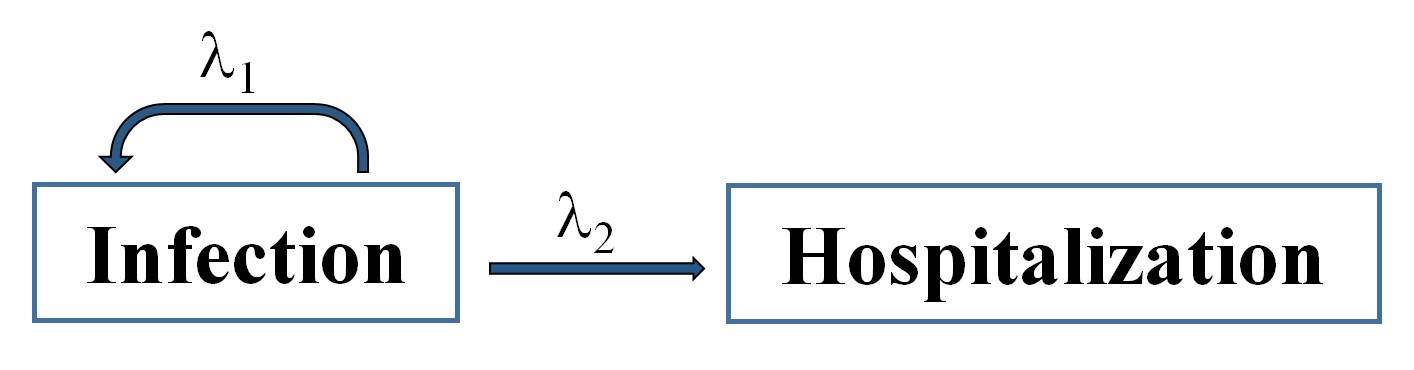}}
	\caption{Transitions diagram: feedback loop from infected to infected and transitions from infected to being in hospital. \label{fig:transitions}}
\end{figure}

 To align our vocabulary to our motivating example on the Covid-19 pandemic and  help the intuition of the reader, we define new increases in $N_1$ as new ``infections'', and the value of $N_1$ at any point as {``number of infections''}. Notice that in our motivating example, $N_1$ is not the number of infections in the population, but the registered number of infections. But to ease the reading, we use the short-hand ``infections''. In the same way, we will take advantage of the intuition of our motivating example and call increases in $N_2$ for ``hospitalizations''. Another application of our methodology is to let $N_1$ be the number of observed  hospitalized and $N_2$ be the number of deaths in hospital. Modelling  the number of deaths in hospital is an important ingredient when understanding the dynamics of a developing pandemic.


\subsection{Modelling the feedback loop of the vaguely defined exposure measure $N_1$} \label{sec:mod1}%
We observe $n$ individuals in the interval $(0, T]$. Let $N_1(t)$ count the number of persons getting infected in the interval $(0,t]$, for $0< t \leq T$. Also, we denote {by} $N_1(\d t)$ the number of persons infected in the interval $(t, t+\d t]$, then $N_1(t)=\int_0^t N_1(\d s)$, for $t \in (0,T]$. 
Furthermore, we write $N_1(\d t,\d s)$ for the number of pairs of persons where person 1 was infected in $(s,s+\d s]$ and person 2 was infected by person 1 in $(t,t+\d t]$, and $N_1(t, \d s)$ for the number of persons infected in $(s,t]$ by a person that was infected in $(s,s+\d s]$, with $s < t \leq T$. 
We assume that we observe the process $N_1(t)$ but not $N_1(\d t,\d s)$ or $N_1(t,\d s)$, and define a model for $N_1$ based on Hawkes{-}type processes.  

We define ${\mathcal F}_1(t)$ as the $\sigma$-field generated by $\{N_1(s): s<t\}$, and
\[
\lambda_1(t)=\lim_{\d t \rightarrow 0} \frac{E[N_1(t+\d t)-N_1(t) \mid {\mathcal F}_1(t)]}{ \d t}, \  0 <t \leq T.
\]
One could allow that several persons are infected at the same time point. In a Hawkes model this is excluded. For simplicity of discussions we also do not allow this. 
 We will only deviate from this assumption when we will introduce a bootstrap procedure for uncertainty quantification of predictions
in Section \ref{sec:uncertainty}. There, a factor $\gamma$ is introduced which corrects for an empirically observed much larger variance than expected from a Hawkes model. Otherwise, in case that the assumption hold, we have that
 $\lambda_1(t)\d t=\E [N_1(\d t)\mid {\mathcal F}_1(t)]$ is  the probability that one person is infected in the interval $(t,t+\d t]$ given ${\mathcal F}_1(t)$. 

For the dynamics of $N_1(t)$ we assume that
\begin{equation}\label{eq:haw1}
	\lambda_1(t)\d t=\left(\int_0^{t^-} \mu_1(t/T,t-s)N_1(\d s)\right)\d t + n \rho_1(t)\d t ,
\end{equation}
where $\rho_1$ and $\mu_1$ are some unknown functions. We assume that $\mu_1(t/T,v)=0$, for $v<0$, $v>D_1$ and that $\rho_1(t)=0$, for $t>D_1^\prime$, where $D_1$ and $D_1^\prime$ are some fixed constants.  The role of $D_1$ and $D_1^\prime$ will be discussed in Remark 3 at the end of Section \ref  {sec:algo1}.

According to equation \eqref{eq:haw1}, $N_1(t)$ is a locally stationary Hawkes process \citep{Mammen:Muller:23}.  We have that, with $0 \leq s <t \leq T$,  $\mu_1({t}/{T}, t-s)\d t$ is the probability that a person infected at $s$ infects a person in $(t,t+ \d t]$. Hence, we are interested in the estimation of $\mu_1(t/T,v)$ for interior points $t$ of the interval $(0,T)$. 

A key step in our paper is, therefore, to estimate the two-dimensional transition intensity $\mu_1(t/T,w)$, for $0<t<T$ and $0 < w < D_1$.  Ideally, to achieve this, we have observations of the {two-dimensional} process $N_1(t,\d s)$. However, in our motivating problem we only observe the {one-dimensional} process $N_1(t)$. In Section \ref{sec:mu_1} we describe the estimation of $\mu_1(t/T,w)$ in the case we had observed $N_1(\d t,\d s)$, for all $s < t \leq T$. Afterwards we propose an algorithm which is able to estimate $N_{1}(\d t,\d s)$ from the observations of $N_1(t)$. 

\begin{remark}
In Hawkes models, the function $\rho_1$ is used to model infections by individuals that were infected before the starting point 0 of the study and  by additional infected persons who enter the population from outside. We are interested in periods of the {epidemic} where {it} is fully developed and where the occurrences  of additional infections from outside are negligible.  Therefore, we assume that $\rho_1(t)$ is equal to zero after an initial period. We consider  $\rho_1$ as a  nuisance function and it is only used to model the {epidemic} during the starting time points of the study, as well as in our asymptotic developments. Knowledge and estimation of $\rho_1$ is not needed in our approach. Our study could be also extended to epidemics where infected persons enter from outside, and where $\rho_1(t)$ is strictly positive for the whole period. From a theoretical point of view this would simplify the development of statistical approaches and their mathematical analysis, see \cite{Mammen:Muller:23}.

The function $\rho_1$ could be chosen for instance as $\rho_1(t)=\int_{-\infty}^0 \mu_1({t}/{T}, t-v)\nu_1(\d v)$, where $n \nu_1(t)$ are the values of $N_1(t)$ in the past $t <0$.
If we condition on the past, we can treat $\rho_1$ in this example as if it is a deterministic function. Note that in this case we see that $\rho_1(t)=0$, for $t>D_1$ because $\mu_1(t/T,w)=0$ for $w >D_1$.  In the following we will make this assumption but our discussions  could be easily generalized to arbitrary specifications of $\rho_1$.  In our data example and our simulations we {use} discretised data and {choose} $\rho_1$ depending on the value of infected persons at the first day. This lead to an inaccurate estimate of $\mu_1$ only at a very small starting period as can be seen {later} in Figure \ref{fig:RoI}, where estimates of $\mu_1$ are shown for different starting dates.

\end{remark}


\subsection{Modelling the relationship between the vaguely defined process, $N_1$, and the events of interest, $N_2$}\label{sec:mod2} 

In this section we build a model for the relationship between the infection process $N_1$ and the hospitalization process $N_2$. We observe $n$ individuals in the interval $(0, T]$. Let $N_2(t)$ count the number of persons hospitalized in the interval $(0,t]$,  and $N_2(\d t)$ the number of  hospitalized in the interval $(t, t+\d t]$. Then $N_2(t)=\int_0^t N_2(\d u)$, for  $0< t \leq T$.  
Furthermore, we write $N_2(\d t,\d s)$ for the number of persons hospitalized in $(t,t+\d t]$ that had been infected in $(s,s+\d s]$, and $N_2(t, \d s)$ for the number of persons infected  in $(s, s+\d s]$ that enter the hospital in the interval $(s,t]$, $s<t \leq T $. We consider the situation where one observes the process $N_1(t)$ and $N_2(t)$ for the whole interval $(0, T]${,  but}  $N_2(\d t,\d s)$ is not observed. 

Let us  define
\[
\lambda_2(t)=\lim_{\d t \rightarrow 0} \frac{E[N_2(t+\d t)-N_2(t) \mid {\mathcal F}_1(t)]}{ \d t}, \  0 <t \leq T,
\]
where ${\mathcal F}_1(t)$ is the $\sigma$-field generated by $\{N_1(s): s<t\}$, with $N_1$ given in Section \ref{sec:mod1}. Then $\lambda_2(t)\d t=\E [N_2(\d t)\mid {\mathcal F}_1(t)]$ is  the probability that one person is hospitalized in the interval $(t,t+\d t]$ given ${\mathcal F}_1(t)$. 
We assume that
\begin{equation}\label{eq:haw2}
	\lambda_2(t)\d t=\left(\int_0^{t^-} \mu_2(t/T,t-s) N_1(\d s)\right)\d t + n \rho_2(t)\d t,
\end{equation}
with $\rho_2$ and $\mu_2$ some unknown functions such that $\mu_2(t/T,v)=0$, for $v<0$ and $v>D_2$, and $\rho_2(t)=0$ for $t>D^\prime_2$, with $D_2, D^\prime_2$ some fixed constants, and where,  for a person, who was infected at $s$,   $\mu_2({t}/{T}, t-s)\d t$ is the probability that this person is hospitalized in $(t,t+\d t]$.

\begin{remark}
From equation \eqref{eq:haw2}, $N_2(t)$ would be a Hawkes process if we would replace ${\mathcal F}_1(t)$ by the $\sigma$-field generated by $\{(N_1(s),N_2(s)): s<t\}$. With this modification the assumption \eqref{eq:haw2} is too restrictive. Typically $N_2(s)$ contains information about the future development of $N_2$. Consider {for instance,} in a thought experiment, the unrealistic case that at a certain date all infected individuals are hospitalized. Then it can be expected that the number of additional hospitalizations in the next days will be low. Another case would be periods where the hospitals are overcrowded and have difficulties to admit more patients. Fitting  model  \eqref{eq:haw2} can be used to predict future hospitalizations {and the prediction would depend only} on the history of $N_1$. We conjecture that in most settings neglecting the history of $N_2$ will not lead to a large loss of accuracy, at least if $n$ is large, $N_2(t)$ is much smaller than $N_1(t)$, and the hospitals are not overcrowded. Furthermore, it would be very complex to model future developments of $N_2$ using the history of $N_1$ and $N_2$.  
\end{remark}

\section{Estimation of the two-dimensional transition intensity}\label{sec:estim} 

 In this section we propose estimators for the two-dimensional transition intensities, $\mu_1$ and $\mu_2$, in the models described above. First, in Section \ref{sec:mu_1}, we propose an estimator for the infection rate, $\mu_1$ that works when we have individual follow up and thus, information is registered of the time since one person is infected until he/she causes a new infection.  In the following, we will refer to this situation with the term ``full information''. Then, in Section \ref{sec:algo1}, we present an algorithm able to generate information of the time-from-infected from available data that does not contain this information. Similarly, in Section \ref{sec:mu_2}, we define an algorithm to  estimate the hospitalization rate, $\mu_2$, when information about the time length since infection to hospitalization is not available. 

\subsection{Estimating the transition intensity from infected to infected with full information}\label{sec:mu_1}

With full information, we observe the process $N_1(t, \d s)$, for $s<t \leq T$.
For $s$ fixed, $N_{1}(t,\d s)$ is a counting process with respect to a (right continuous, increasing and complete) filtration $\mathcal{F}_1(t)$, with $t \in (0, T)$. We are interested in estimating the two-dimensional intensity, $\mu_1(t/T,t-s)$,  with no restriction on its functional form.
We consider the two-dimensional local linear estimator given by:

\begin{eqnarray}\label{eq:alpha1}
	&&\widehat{\mu}_1(t/T,t-s)\\
	\nonumber &&=
	\frac{ \int_{0\leq v < u \leq T} {\rm C}_1(s,t,v,u) \ K_{1,b_1}\left(\frac{t-u}{T}\right) K_{2,b_2}(t-s-(u-v)) N_{1}(\d u,\d v)}
	{ \int_{0\leq v < u \leq T}{\rm C}_1(s,t,v,u) \ K_{1,b_1}\left(\frac{t-u}{T}\right) K_{2,b_2}(t-s-(u-v)) N_{1}(\d v) \ \d u},
\end{eqnarray}
with  $N_{1}(\d u,\d v)$  and  $N_{1}(\d v)$ as defined previously.

The estimator consists of a ratio of a local linear estimator of occurrences and a local linear estimator of exposure. This is similar to marker dependent kernel hazard estimation in survival analysis. See  \cite{Nielsen:Linton:95} for the local constant case, \cite{Nielsen:98} for the local linear case, and \cite{Gamiz:etal:13} for the problem of bandwidth selection in this context. 
Readers are referred to Appendix \ref{ap:estim} for a comprehensive understanding of the local linear estimator presented in this section, as well as a detailed explanation of the notation used in equation \eqref{eq:alpha1}. 

For the choice $C_1(s,t,v,u) \equiv 1$ one has the mathematically simpler case of local constant estimation. We now explain that then the right hand side of  \eqref{eq:alpha1} is a ratio of a local constant estimator of occurrences and a local constant estimator of exposure. We do this for the special case that $K_{1,b} $ is a uniform kernel with bandwidth $b$. Define $S_1$ as  the number of pairs $(U,V)$ such that $|U-V-(t-s)| \leq b_2$, $|U -t| \leq b_1$ and two individuals exist, where the first was infected at $U$ by a person infected at $V$. Further denote by $S_2$ the integral over $u$ such that $|u-V-(t-s)| \leq b_2$, $|u -t| \leq b_1$ and an individual exists that was infected at $V$. Then the ratio on the right hand side of (3) is given as $(S_1/(4b_1b_2))/(S_2/(4b_1b_2)) = S_1/S_2$. 
This can be interpreted as  ratio of an estimator of occurrences and an estimator of exposure.
In our implementations we used other kernels and a choice of $C_1(s,t,v,u)$ leading to local linear estimation, see Appendix \ref{ap:estim}. This is motivated by smoothing theory that recommends use of local linear instead of local constant smoothing.


\subsection{An iterative estimation scheme approximating the infeasible estimator of the last subsection}

\label{sec:algo1} 

We now describe an iterative estimation scheme for the estimation of $\mu_1$, {in the real case scenario where  we do not observe the process $N_1(t,\d s)$ but only $N_1(t)$. 
Starting with an initial guess, $\widehat{\mu}_1^{(0)}$, of $\mu_1$, the} $r$th iteration of the algorithm consists of two steps. In the first step
we construct a two-dimensional process $\widehat N_1^{(r)}(t,\d s)$ that approximates $N_1(t,\d s)$. This is done by using the estimator $\widehat{\mu}_1^{(r-1)}$ from  the previous iteration  and  by using the  observed process $N_1(t)$ {as follows}:
\begin{equation}\label{eq:alpha1a}\widehat N_1^{(r)} (\d u, \d v)= \frac { \widehat \mu_1^{(r-1)} (u/T, u-v) }
	{\int_0^{u^-}\widehat \mu_1^{(r-1)} (u/T, u-w)  N_1(\d w)} N_1(\d v) N_1(\d u). \end{equation}
 For a motivation of the iteration \eqref{eq:alpha1a} note that for an interval $I=[a,b]$ and $u > b$, given that an infection takes place at $u$, the probability that this infection was caused by an individual getting infected in the interval $I$ is equal to  $ \int_{v \in I}  \mu_1 (u/T, u-v) N_1(\d v)/
	\int_0^{u^-} \mu_1 (u/T, u-w)  N_1(\d w)  $. This shows that approximately
 $$\int_{v \in I} N_1(\d u, \d v) \approx \int_{v \in I} \frac{ \mu_1 (u/T, u-v)}	{\int_0^{u^-} \mu_1 (u/T, u-w)  N_1(\d w)} N_1(\d v) N_1(\d u).$$
 This motivates the update \eqref{eq:alpha1a}.
In the second step, the {approximation} \eqref{eq:alpha1a} of   $ N_1 (\d u, \d v)$ is plugged into \eqref{eq:alpha1}{, providing the} following update of $\widehat{\mu}_1^{(r-1)}$:

\begin{eqnarray}\label{eq:alpha1b}
	&&\widehat{\mu}_1^{{(r)}}(t/T,t-s)\\
	\nonumber &&= \frac{ \int_{0\leq v < u \leq T} {\rm C}_1(s,t,v,u) \ K_{1,b_1}\left(\frac{t-u}{T}\right) K_{2,b_2}(t-s-(u-v)) \widehat N_1^{(r)}(\d u,\d v)}
	{ \int_{0\leq v < u \leq T}{\rm C}_1(s,t,v,u) \ K_{1,b_1}\left(\frac{t-u}{T}\right) K_{2,b_2}(t-s-(u-v)) N_1(\d v) \ \d u}.
\end{eqnarray}
In our simulations and in our data application we observed that this iteration scheme converges very fast. 

In our asymptotic analysis we will assume that  $ \widehat \mu_1 (t/T, t-s)$ is an estimator that fulfils the following equation: 
\begin{eqnarray}
	\nonumber  &&\widehat \mu_1(t/T, t-s) =\int_{0 \leq v < u \leq T}{\rm C}_1(s,t,v,u) K_{1,b_1} \left ( \frac {t-u} T\right ) 
	\\ \nonumber  && \qquad \times K_{2,b_2} (t-s - (u-v)) \frac { \widehat \mu_1 (u/T, u-v) }
	{\int_0^{u^-}\widehat \mu_1(u/T, u-w)  N_1(\d w)} N_1(\d v) N_1(\d u)  \\ && \qquad \times \frac 1 {  \int_{0\leq v < u \leq T} {\rm C}_1(s,t,v,u) K_{1,b_1} \left ( \frac {t-u} T\right ) K_{2,b_2} (t-s - (u-v))
		N_1(\d v) \d u} \nonumber\\ && \qquad + o_P(1). \label{estdefeq}
\end{eqnarray}
A possible choice for an estimator that fulfils \eqref{estdefeq} would be the just described iterative estimator where the iteration is stopped if the difference between the left-hand side and the right-hand side of \eqref{eq:alpha1b} is below a threshold of order $o(1)$. 
In Appendix \ref{sec:theo1}, we will argue that solutions $ \widehat \mu_1 (t/T, t-s)$  of equation \eqref{estdefeq} are consistent  estimators.

\begin{remark} 
	In our theoretical model, we assume that $\mu_1(t/T, v) = 0$ for $v > D_1$ meaning that an infected individual can only transmit the infection to others within a time period no longer than $D_1$. In our numerical applications, we do not assign a specific value to $D_1$.
	Due to the "vague exposure" context, our algorithm extracts information about the number of individuals infected either directly or indirectly by others. As a result, we obtain that $\widehat{\mu}_1(t/T, t - s) > 0$ for $0 < s < t \leq T$. Thus, $\widehat{\mu}_1(\cdot/T, v)$ may take non-zero values for $v \in (0, T]$. 
	On the other hand, in our approach {$\rho_1$} is treated as a nuisance function then we do not specify any value for $D_1'$. 
	
\end{remark}

\subsection{Estimating the transition intensity from infected to hospitalized}\label{sec:mu_2} %

We now describe an iterative algorithm to estimate the transition intensity $\mu_2$. The procedure uses a similar approach as for the estimation of $\mu_1$, discussed in the last section.  
Starting with an initial guess, $\widehat{\mu}_2^{(0)}$, of $\mu_2$, the $r$th iteration of the algorithm consists of two steps. In the first step we construct a two-dimensional process $\widehat N_2^{(r)}(t,\d s)$ that approximates $N_2(t,\d s)$. This is done by using the estimator $\widehat{\mu}_2^{(r-1)}$ from  the previous iteration{,  and } the  observed processes $N_1(t)$ and $N_2(t)$ {as follows (compare with equation \eqref{eq:alpha1a}):}
\begin{equation*}\widehat N_2^{(r)} (\d u, \d v)= \frac { \widehat \mu_2^{(r-1)} (u/T, u-v) }
	{\int_0^{u^-}\widehat \mu_2^{(r-1)} (u/T, u-w)  N_1(\d w)} N_1(\d v) N_2(\d u). 
\end{equation*}
In the second step, we proceed similarly to \eqref{eq:alpha1b} to get the update:
\begin{eqnarray*}
	&&\widehat{\mu}_2^{{(r)}}(t/T,t-s)\\
	&&= \frac{ \int_{0\leq v < u \leq T} {\rm C}_1(s,t,v,u) \ K_{1,b_1}\left(\frac{t-u}{T}\right) K_{2,b_2}(t-s-(u-v)) \widehat N_2^{(r)}(\d u,\d v)}
	{ \int_{0\leq v <u \leq T}{\rm C}_1(s,t,v,u) \ K_{1,b_1}\left(\frac{t-u}{T}\right) K_{2,b_2}(t-s-(u-v)) N_1(\d v) \ \d u}.
\end{eqnarray*}
If the algorithm converges{, then} the limit $\widehat{\mu}_2$ of $\widehat{\mu}_2^{(r)}$ {will fulfil the equation:} 
\begin{eqnarray*}
	\nonumber  &&\widehat \mu_2(t/T, t-s) \\
	&&=\int_{0 \leq v < u \leq T}{\rm C}_1(s,t,v,u) K_{1,b_1} \left ( \frac {t-u} T\right ) K_{2,b_2} (t-s - (u-v))
	\\  && \qquad \times \frac { \widehat \mu_2(u/T, u-v) }
	{\int_0^{u^-}\widehat \mu_2(u/T, u-w)  N_1(\d w)} N_1(\d v) N_2(\d u)  \\ && \qquad \times \frac 1 {  \int_{0\leq v < u \leq T} {\rm C}_1(s,t,v,u) K_{1,b_1} \left ( \frac {t-u} T\right ) K_{2,b_2} (t-s - (u-v))
		N_1(\d v) \d u}. \nonumber
\end{eqnarray*}
The performance of this estimator can be discussed by using similar arguments as that of $\widehat \mu_1$ in the last section {(compare with equation \eqref{estdefeq})}. In particular this concerns the asymptotic analysis.
}


\section{Principles of forecasting in a dynamic environment}\label{sec:forecast} 

Let us define a variable (denoted later by $C_{t,h}$) indicating at any point in time $t$ whether the future ($t+h$, $h > 0$) will be equal or different to the immediate past. 
There might be periods where little is happening {and the variable might have a tendency to increase/decrease slowly,} and there might be few but very important change-points where interventions are introduced (e.g. a lock down to minimize future infections) and the variable might drop dramatically in a matter of days.

Let $t^*$ be the calendar time for the most recent estimate. From the observed data (until $t^*$) we can estimate the dynamic infection rate as  $\widehat \mu_1 \left(t/T, u\right)$, for $0< u \leq t \leq t^*$. Let us fix the forecast horizon at time $t^*+h$, for an $h>0$. {Then we} need to extrapolate the dynamic infection rate  $\mu_1\left((t^*+s)/T,u\right)$, for $u \geq 0$ and $0< s \leq h$. {Following the above idea, we} assume that the infection rate at the end of the forecasting period is the more recent estimate multiplied by {a number denoted by} $C_{t^*,h}$, and it varies linearly in between{, this is,}
\begin{equation}\label{eq:fore1}
	\widetilde\mu_1\left((t^*+s)/T,u\right)=\widehat\mu_1\left(t^*/T,u\right) \times \left (1+ (C_{t^*,h}-1)\frac{s}{h}\right ),
\end{equation}
for $0\leq s \leq h$ and $u\geq 0$.

One can monitor and forecast with $C_{t^*,h}$ equal to one. And one can use {knowledge from other sources}  (e.g. knowledge of intervention) to forecast with a $C_{t^*,h}$ different from  one. A case study of Covid-19 in France will be {described in the next section. In that case} most of the dynamic is happening from infection to infection and one can therefore concentrate {external} knowledge around this transition, while keeping the transition from infection to hospitalization unaffected by {external} knowledge. In that situation {the dynamic statistical methodology  introduced} in this paper can do the job via a surprisingly simple data collection, reflecting what can be considered ``available data'' in many countries. Therefore, the only thing left to monitor and forecast well is to have a dynamic point of view {of the variable} $C_{t,h}$. That work needs expert advice specific to countries or local regions within countries. This important work cannot be dismissed via a statistical analysis based on simplified available data. But it is important that the experts involved should only consider forecasting and understand the {variable} $C_{t,h}$, rather than being responsible for a full statistical model.  

In our data example we consider the so-called reproduction number, or simply $R$-number, that is defined as the expected number of new infections caused by an infectious individual \citep{Fraser:07}. Estimates of the actual Covid-19 reproduction numbers have been publicly reported in most countries. In the next section we will {suggest} choosing $C_{t^*,h}$ such that the reproduction number {corresponding to} $\widetilde\mu_1$, {that is} given by 
$$\int \widetilde\mu_1\left((t^*+s)/T,u\right) \d u = \int \widehat\mu_1\left(t^*/T,u\right) \d u
\times \left (1+ (C_{t^*,h}-1)\frac{s}{h}\right ), $$
is equal to the reported {and externally} estimated reproduction number for $s=h$. To understand this, first note that $\widetilde{\mu}\left((t^*+h)/T,u\right)=C_{t^*,h}\times \widehat\mu_1\left(t^*/T,u\right)$. This implies that
$$\int \widetilde\mu_1\left((t^*+h)/T,u\right) \d u=C_{t^*,h}\times\int_{0}^{t^*}\!\!\widehat\mu_1\left(t^*/T,u\right) \d u. $$
Here, we are using that $\widehat{\mu}(t/T,u)$ is either not defined or equal to zero for $u>t$. The integral on the right-hand gives the total number of infected in the interval $(0, t^*]$. Therefore, $C_{t^*,h}$ represents the expected number of infections caused by each person infected during this time period, which corresponds to the common definition of the reproduction number, according to \cite{Fraser:07}.


\section{Estimating with available data. The Covid-19 case of France}\label{sec:application} 

 We consider French Covid-19 data on the reported number of daily infections and hospitalized patients during the pandemic.  To model these data we combine the model of the development of observed infected in the population, presented above, with the model linking the observed number of infections to the number of arrivals to the hospital. 

{In the raw data, we observe that the number of reported infections varies significantly depending on the day of the week, with typically fewer cases reported on weekends compared to weekdays. Although these fluctuations may reflect some variation in transmission patterns related to human behaviour, they are more likely the result of reporting delays and slower case confirmation processes that typically occur over the weekend. To prevent this from affecting our predictions, we process the data according to the method described in \cite{Koyama:et:al:21}, before analysing the sequence of daily cases. See Appendix \ref{app:koyama} for more details.}

Using the methodology of Sections \ref{sec:mu_1} and \ref{sec:algo1}, we estimate the infection rate, $\mu_1$, describing the transition loop from infected to infected, to get the transition rates shown in Figure \ref{fig:RoI}. And,  {with} the methodology in Section \ref{sec:mu_2}, we estimate the hospitalization rate, $\mu_2$, describing the transition from infection to hospitalization, deriving the transition rates shown in Figure \ref{fig:RoH}. 
The numerical calculation of the continuous integrals (involved in the expressions of the estimators) is performed via discrete approximations to these integrals based on our available daily observations, similarly to \cite{Gamiz:etal:13}. Furthermore, optimal bandwidths have been estimated for simplicity using the cross-validation method described in \cite{Gamiz:etal:13}. This method worked well for this application, otherwise other methods with better properties such as the Do-validation method of \cite{Gamiz:etal:16} could be used, with a higher computational cost (see also \cite{Mammen:etal:11}).


\begin{figure}[htb]
	\centering
	\makebox{\includegraphics[scale=0.5]{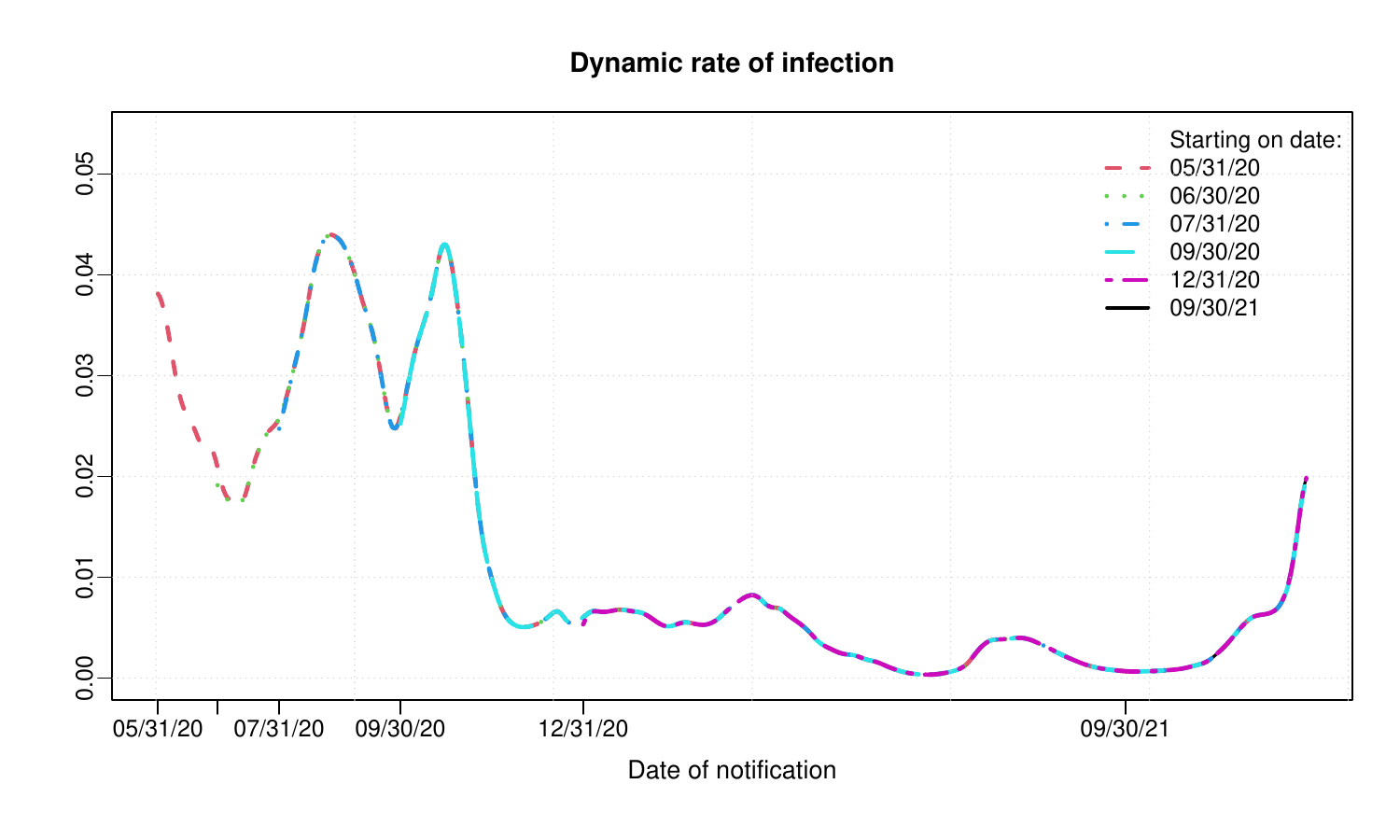}}
	\caption{Dynamic estimation of the rate of infection. Each line on the plot consists of the estimated intensity of the infection process that initiated at the date indicated in the legend, and represented in terms of $\alpha_1(s,t)$ ($t >s$) at six calendar times $s$, taken from May 2020 to December 2021. \label{fig:RoI}}
\end{figure}

\begin{figure}[htb]
	\centering
	\makebox{\includegraphics[scale=0.5]{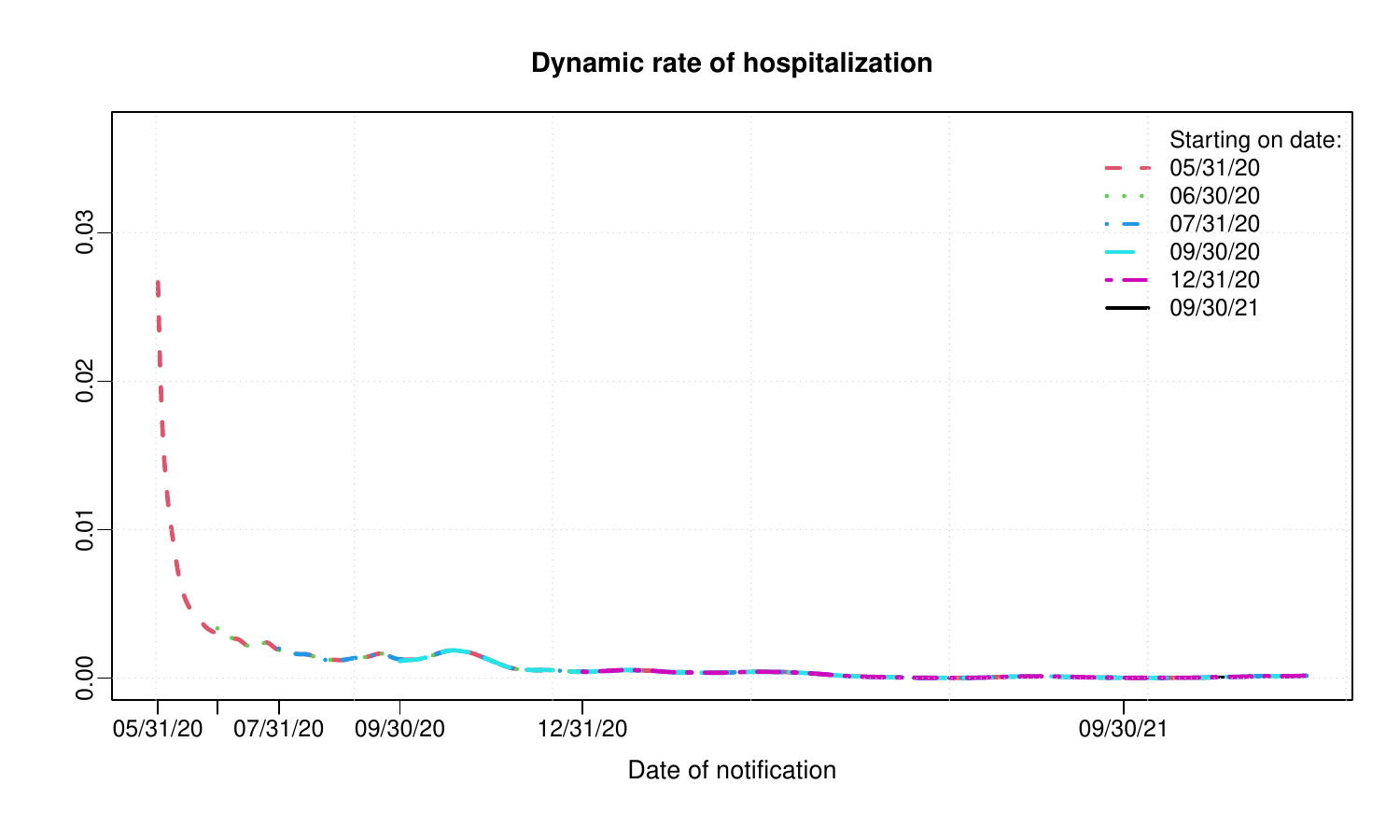}}
	\caption{ Dynamic estimation of the rate of hospitalization. Each line on the plot consists of the estimated  hospitalization rate for individuals infected on the date indicated in the legend, and represented in terms of $\alpha_2(s,t)$ ($t >s$) at six calendar times $s$, taken from May 2020 to December 2021. \label{fig:RoH}}
\end{figure}

Figures \ref{fig:RoI} and \ref{fig:RoH} are designed to visualize the dynamics of infections and hospitalizations, respectively, during the French Covid-19 pandemic. To aid in understanding the situation, we have chosen to represent these transition rates from a perspective different from that provided by the functions $\mu_1(t/T,\cdot)$ and $\mu_2(t/T,\cdot)$ for infections and hospitalizations, respectively, which is the notation that we mainly use in the paper. 
Specifically, for $s$ fixed, we denote by $\alpha_1(s,t)$ the infection rate at time $t$ of a process that initiated at time $s$. Thus, we have $\alpha_1(s,t) = \mu_1(t/T, t-s)$, for $0 < s < t \leq T$. Similarly, we denote  by $\alpha_2(s,t)$ the hospitalization rate at time $t$ for an individual infected at time $s$, and then we have $\alpha_2(s, t) = \mu_2(t/T, t-s)$, $0<s<t\leq T$.
With this notation, Figure \ref{fig:RoI} shows the estimates of $\alpha_1(s,t)$ at six different starting dates $s$. Specifically, the red dashed curve in Figure \ref{fig:RoI} represents the rate of the infection process starting on 31st of May, 2020. For instance, on 31st of July, 2020, the red curve indicates that new infections are occurring at a rate of 0.025, directly or indirectly caused by individuals infected on 31st of May, 2020. We can track this infection line (direct and indirect cases) up to the most recent date in our observation window, which ends on 31st of December, 2021. 
The same end date applies to all six curves represented in the figure. We believe that Figure \ref{fig:RoI} provides a clear visual representation of the infection process for the entire population. The curves initially show different behaviours near the starting point, but after some time, they converge and overlap. This suggests that after a certain period, following the start of the infection process, new infections are less likely to be directly caused by individuals who tested positive on the reference date. Instead, the new infections are more likely due to recently infected individuals, although these may still be indirectly linked to the original cases. 
Similarly, Figure \ref{fig:RoH} displays estimations of $\alpha_2$. In particular, the red dashed curve in Figure \ref{fig:RoH} represents the estimated transition rate from infection to hospitalization for a person infected on 31st of May, 2020. As seen in the plot, the likelihood of being hospitalized decreases rapidly in the days immediately after infection and then stabilizes near zero. Furthermore, the plots show that the severity of Covid-19 was much greater during the early stages (e.g. May 2020) compared to the later stages (e.g. December 2021).

  A steep decline in the number of cases in both May and November 2020 is noticeable in Figure \ref{fig:RoI}, which can be explained as a result of the severe social restrictions that were imposed in France at those moments in time. Figure \ref{fig:RoH} shows a perhaps surprising issue: a sharp descending trend of the hospitalization rates when are seen as a function of the date of onset. That is, about 5\% of new cases is estimated to be hospitalized on the 31st of May, 2020; after the 30th of June same year, there will be less than 1\%;  and below 0.5\% in 2021. This might suggest a slowdown in the speed of arrivals to hospital with time, which may be strange given the dramatic increase in number of cases in the last period, but can be explained due to several factors. One could be the effect of the vaccines and other the successive variants of the virus, which were spreading faster but were less lethal.


\section{  Principles of forecasting. The Covid-19 case of France}\label{sec:fore_France} 

 In this section we combine the estimation methods of Section \ref{sec:application} with the principles of forecasting described in Section  \ref{sec:forecast}. This section therefore provides us with a full forecasting methodology of number of  infections and  hospitalizations. To demonstrate our methodology, we outline the process of forecasting new infections and hospitalizations for October 2020, based on French Covid-19 data until 30th of September. Additional examples are provided in the Supplementary Material, including the French Covid-19 case using weekly data, and a preliminary analysis of the UK case.

\subsection{ Forecasting the infection rate  }\label{sec:forecast_Inf} 

It is not an easy task to forecast the infection rate, as relying on immediate past data might not be sufficient to accurately forecast the immediate future. The first challenge we face {when forecasting is the choice of $C_{t,h}$}, defined in Section \ref{sec:forecast}. Selecting an appropriate {value of} $C_{t,h}$ is essential and it might require additional information. 



Figure \ref{fig:Ct} shows a graph of the optimal 
{values of the variable  $C_{t,h}$ (defined in Section \ref{sec:forecast}) at different calendar times $t$,} for one-week-ahead forecasts ($h=7$ days), in the period May 2020 to January 2022. There are few discontinuities corresponding to few observed inconsistencies on the raw data and for which we have not computed the optimal values.  
Figures \ref{fig:foreInf} and \ref{fig:foreHosp} described later reveal that assuming no change in the infection rate (that means setting  $C_{t,h}=1$) is often not a good choice for forecasting the immediate future.  In other words, relying solely on recent data may not accurately predict the near future.  The point of view taken in this paper is that more information, including expert opinion, is necessary to make a good choice of $C_{t,h}$, at any given {calendar time $t$}.

The Covid-19 data analysed in this paper shows that $C_{t,h}$ is closely related to the much published reproduction number 
\citep{Fraser:07} on how many new infected are caused by one infected individual within a given time period. It is reasonably accurate to assume that if the reported reproduction number, hereafter denoted by $R_t$,} remains constant, it should correspond to $C_{t,h}$ equal to one. If $R_t$ 
 is expected to increase then one would need a $C_{t,h}$ bigger than one, and vice versa if $R_t$  is expected to decrease in the immediate future then one would need $C_{t,h}$ smaller than one. These observations are supported by Figure \ref{fig:Ct}, which shows the observed relationship between the reproduction number~---~published weekly in France on the official website during the Covid-19 pandemic~---~and our optimal $C_{t,h}$ value. 

The value of $C_{t,h}$ plotted on Figure \ref{fig:Ct}, for example on the 7th of September 2020, is the optimal value for predicting new infections in the week ending on that exact day, based on data up to that date. In contrast, according to the official website, the reproduction number reflects the epidemiological situation one week prior to notification. Specifically, the $R_t$ value reported on 14th of September 2020 is calculated using data up to the 7th of September, but  it is published one week later. Therefore, in the graph we plot the $R_t$ series not according to the publication date but against the date one week earlier. Taking this into account, the graph shows that each $C_{t,h}$ is very close to $R_t$ value published seven days later. Furthermore, there is a high correlation between the two indicators.  As a result, we believe that a reasonable good  {choice} of $C_{t,h}$ for forecasting 
could derive from a simple regression of optimal values $C_{t,h}$ with actual and predicted $R_t$ numbers. This is clearly an approach we would recommend during a developing pandemic.  In the next section we illustrate forecasting the period 1--31 October 2020 using French Covid-19 data, with a fixed given $C_{t,h}$ value and, in Section \ref{sec:fore_Rt}, we illustrate forecasting with $C_{t,h}$ based on the reproduction number.  In this paper we do not discuss statistical estimation of the reproduction number. We think of applications of our method where the reproduction number is calculated by experts who may have relevant prior knowledge. But there exist also statistical approaches. E.g. for a Bayesian approach to estimate the time-varying reproduction number  using only
daily incidence data and the
time between symptom onset in a case and their infection, see 
\cite{Nash:et:al:23}.

\begin{figure}[ht]
	\centering
	\makebox{\includegraphics[scale=0.4]{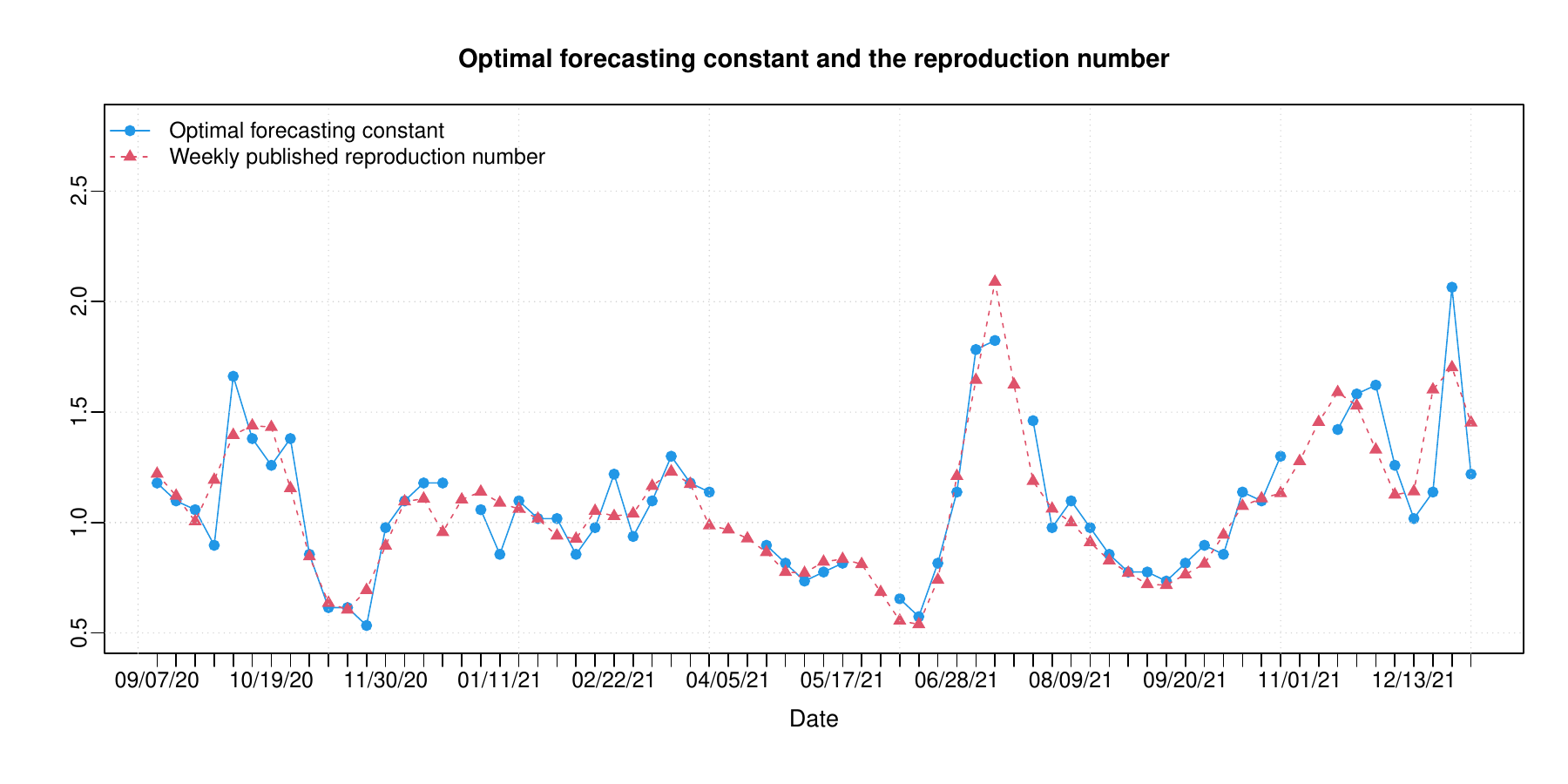}}
	\caption{The optimal estimated values of $C_{t,h}$, at different calendar times $t$, for one-week predictions ($h=7$ days), are plotted along with the
		weekly published reproduction number $R_t$ that reflects the epidemiological situation at time $t$. \label{fig:Ct}}
\end{figure}

\subsection{Forecasting {the period 1--31} October 2020. The Covid-19 case of France}\label{sec:forecast_Oct} 

\begin{figure}[htb]
	\centering
	\makebox{\includegraphics[scale=0.35]{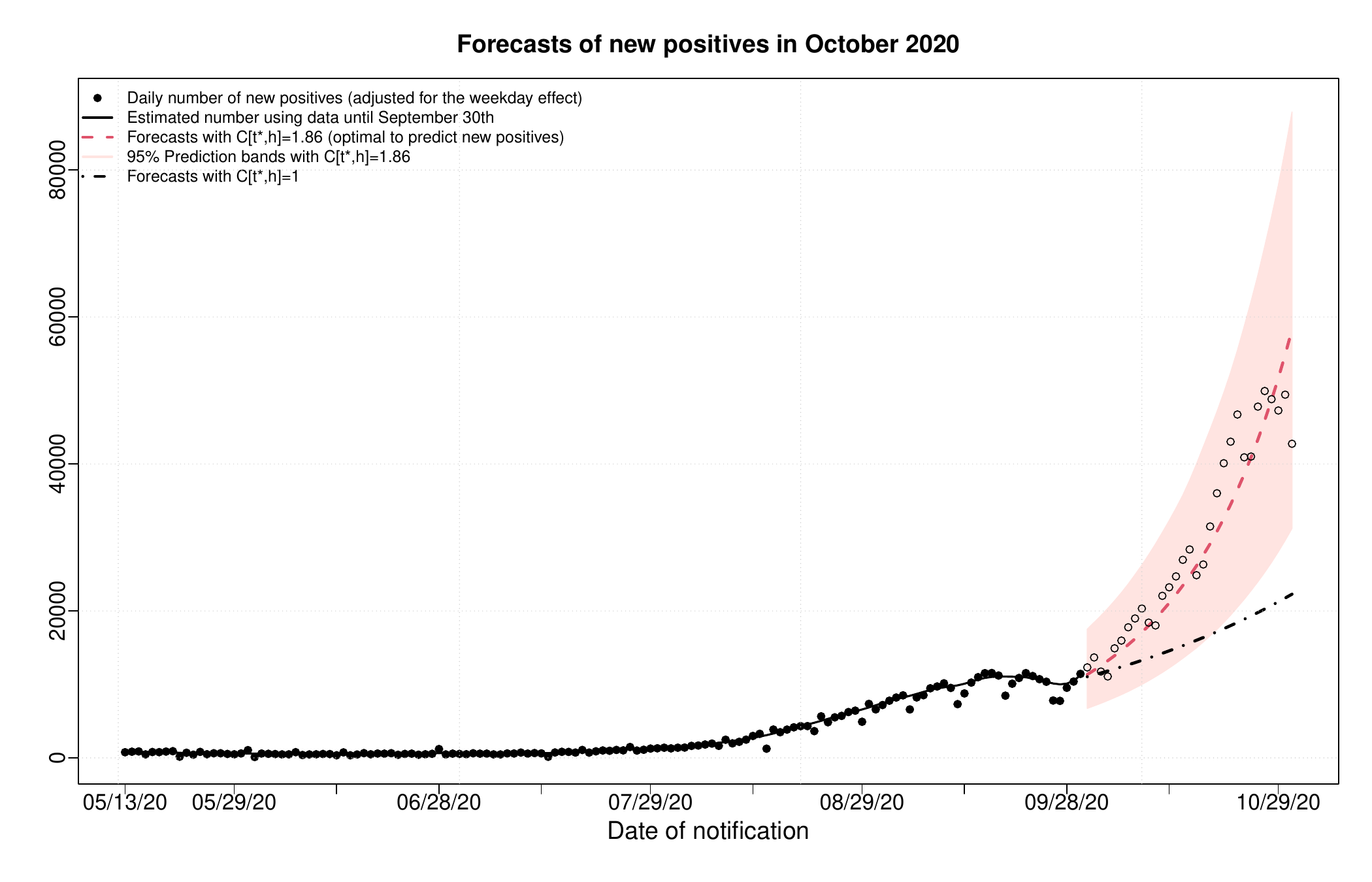}}
	\caption{Number of new positives predicted in the period 1-31 October 2020 using $C_{t^*,h}=1$ (black dotted-dashed line) and using the optimal value $C_{t^*,h}=1.86$ (red dashed line), for $h=31$ days. Actual (adjusted) numbers in October are shown with circles for a reference. \label{fig:foreInf}}
\end{figure}

 Figure \ref{fig:foreInf} provides a graphical representation of the daily reported new infections between the 15th of May and the 31st of October 2020. In this section, our goal is to forecast the number of new infections during October using data available up to the 30th of September. The black dots represent the reported daily numbers up to the 30th of September (used for estimation), while the actual numbers for October are also displayed as circles in the graph, serving as a reference for the forecasts, although they are not used for estimation. As described previously in Section \ref{sec:application}, the data have been adjusted for the weekday effect. Following our notation in previous sections, our most recent estimates are then computed at calendar time $t^*$ being 30th of September. 
{Using} the actual  number of infected reported in October (circles), we can obtain the (infeasible in practice) optimal value of $C_{t^*,h}$ to forecast the infection rate at the end of the prediction period. This gives the value $C_{t^*,h}=1.86$, what means that the rate of infection on 31st of October is $1.86$ times the value of the rate on 30th of  September. Using linear interpolation we get the infection rate for the whole prediction period which allows to obtain the daily new positives in October (red dashed line) that best fit the true values. On the graph we have also shown the forecasts (black dotted-dashed line) for the case of $C_{t^*,h}=1$, this is, assuming that there are no changes in the behaviour of the infection rate in October with respect to what we had at the end of September. 

Note that this assumption seems to be quite wrong when we look at the actual numbers (represented by circles). The uncertainty of the predictions is shown on the graph by representing 95\% prediction bands for the predictions with the optimal $C_{t^*,h}$ value. These bands have been computed using the bootstrap method defined in Section \ref{sec:uncertainty} below.  We have performed an alternative analysis using weekly instead of daily data with similar results that can be seen in the Supplementary Material. 

Figure \ref{fig:foreHosp} presents a similar study as in Figure \ref{fig:foreInf}, but considering the problem of predicting the number of new hospitalizations in October 2020, from historical data until 30th of September. Following the transition diagram in Figure \ref{fig:transitions}, we recognize that forecasting the number of new hospitalizations involves an additional step. Besides extrapolating the infection rate, we also need to extrapolate the hospitalization rate. Here we have assumed that the hospitalization rate at the end of the forecasting period is exactly the hospitalization rate at the end of the estimation period. And, to obtain the optimal $C_{t^*,h}$  value to extrapolate the infection rate, we have proceeded similarly to the previous case, assuming that the rate of infection at the end of the forecasting period is $C_{t^*,h}$ times the value of the infection rate we had estimated for the 30th of September. In this case, we derive infeasible forecasts (red dashed line), using the optimal value of $C_{t^*,h}$ that minimises the error of prediction with respect to the number of hospitalized during the month of October, instead of  the number of infected. And it turns out that the optimal $C_{t^*,h}$ leads us to a rate of infection on 31st of October that is 3.07 times the rate of infection on 30th of September. Again, taking $C_{t^*,h}=1$ in this case would lead to a situation with respect to the number of new hospitalized  (black dotted-dashed line) which is very far from the true situation.  Similar conclusions can be derived from the weekly data analysis in the Supplementary Material, with the optimal value $C_{t^*,h}=2.13$ in this case.

\begin{figure}[htb]
	\centering
	\makebox{\includegraphics[scale=0.35]{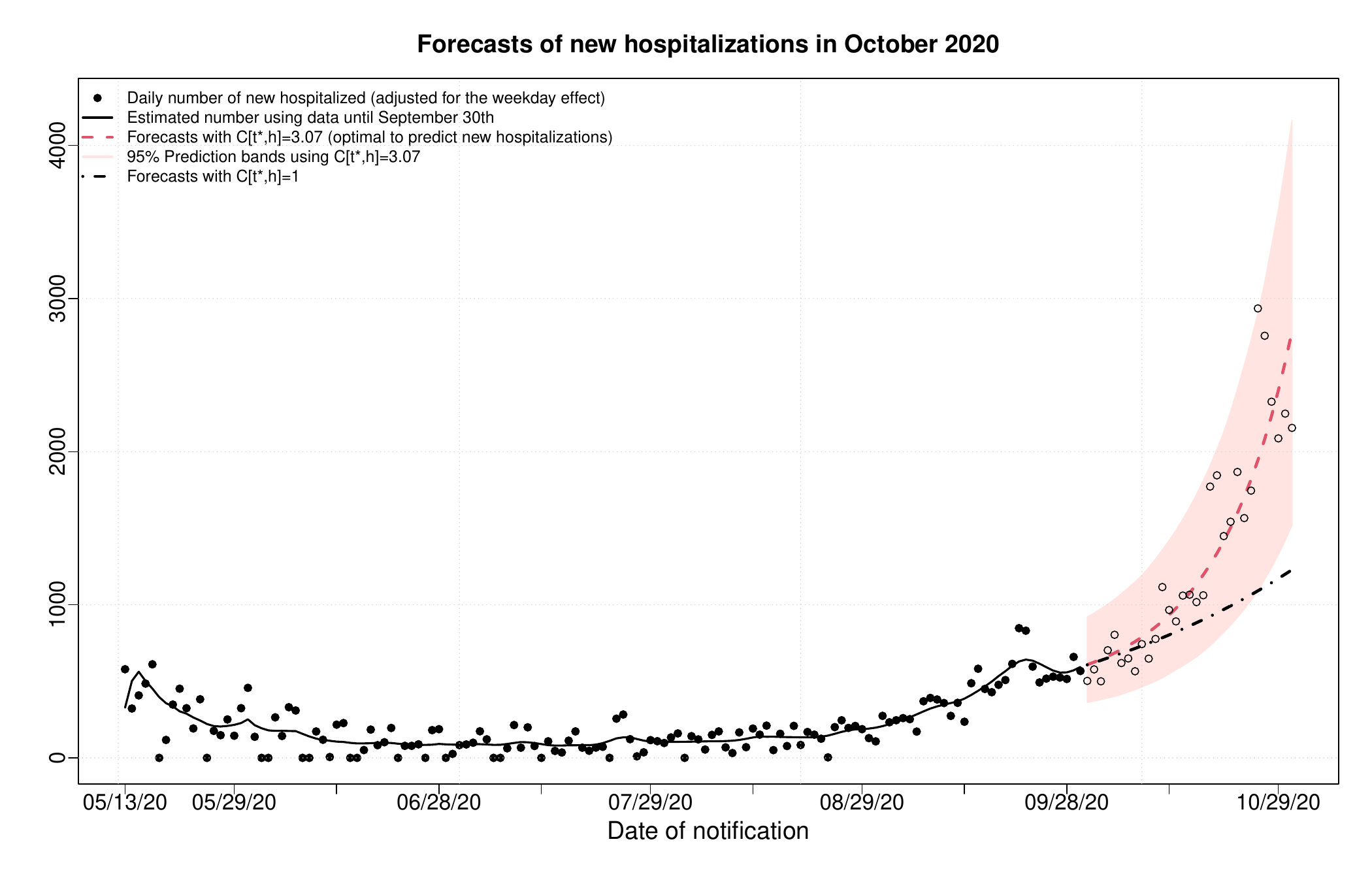}}
	\caption{Number of new hospitalized predicted in the period 1-31 October 2020 using $C_{t^*,h}=1$ (black dotted-dashed line) and using the optimal value $C_{t^*,h}=3.07$ (red dashed line), for $h=31$ days. Actual (adjusted) numbers in October are shown with circles for a reference. \label{fig:foreHosp}}
\end{figure}


\begin{figure}[htb]
\centering
\makebox {\includegraphics[scale=0.35]{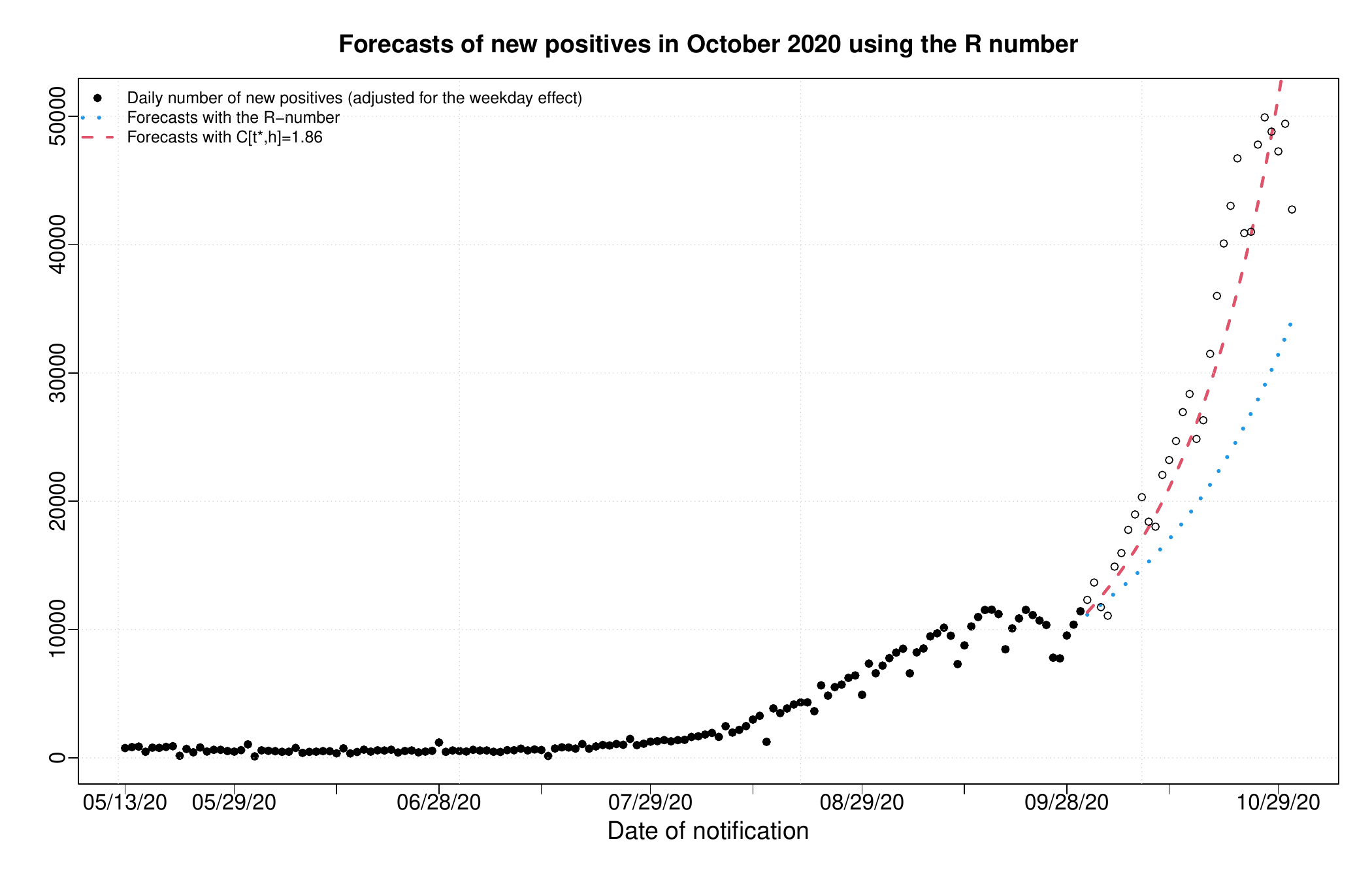}}
\caption{Number of new positives predicted in the period 1-31 October 2020 using the averaged reported reproduction numbers, $\bar R_{t^*}=1.34$ (blue dotted line) and using the optimal value $C_{t^*,h}=1.86$ (red dashed line), for $h=31$ days. \label{fig:foreInf_R}}
\end{figure}

{
\subsection{Uncertainty  quantifications of predictions}\label{sec:uncertainty}

We propose a novel bootstrap method to quantify uncertainty by constructing prediction bands for a fixed given $C_{t,h}$ value. Our bootstrap method assumes  a more general model than the Hawkes model, which is able to account for the overdispersion observed in the dataset. The method has been illustrated with French Covid-19 data on infections and hospitalizations in Figures \ref{fig:foreInf} and \ref{fig:foreHosp} above. Below we provide an outline of the method. 

Assume we observe $N_{1,0}, N_{1,1}, \ldots, N_{1,T}$, where $N_{1,i}$ is the number of infections at the $i$th day ($i=1,2,\ldots, T$). We can assess potential overdispersion in the data by evaluating the overdispersion factor given by
\begin{equation}\label{eq:gamma}
\gamma=(1/T)\sum_{i=1}^T\frac{(N_{1,i}-\lambda_{1,i})^2}{\lambda_{1,i}},
\end{equation}
which compares the empirical variance to the expected variance under the Hawkes model assumption. Here $\lambda_{1,i}$ is the expected number of infections at the $i$th day, given $N_{1,0},\ldots, N_{1,i-1}$. 

For the data presented before, the factor in \eqref{eq:gamma} is estimated as $\gamma=60.0302$, indicating substantial overdispersion relative to the Hawkes model. To account for this, we have quantified the uncertainty of our predictions for the number of infections using a bootstrap method that consists of the following steps:

	\begin{enumerate}
		\item[1.]  Take $k>\gamma $.
		\item[2.] Define $\beta= (\gamma-1)/(k(k-1))$, and $\alpha= 1-k \beta$.
		\item[3.] Given $N_{1,0}, N^*_{1,1}, \ldots, N^*_{1,i-1}$, generate a bootstrap sample with  $N^*_{1,\alpha,i} \longrightarrow Pois(\alpha \lambda^*_{1,i})$, and  $N^*_{1,\beta,1} \longrightarrow Pois(\beta \lambda^*_{1,i})$, 
		with $\lambda^*_{1,i}=N_{1,0} \hat{\mu}_1(i,i)+ N^*_{1,1} \hat{\mu}_1(i,i-1)+\ldots +N^*_{1,i-1} \hat{\mu}_1(i,1)$, and define 	$N^*_{1,i}=N^*_{1,\alpha,i}+ k \ N^*_{1,\beta,i}$.
	\end{enumerate}
Repeating step 3 above a large number of times, we are able to compute 95\% prediction limits by evaluating the 2.5\% and 97.5\% quantiles of the bootstrap samples at each day in the forecasting horizon. This resampling scheme mimics the possibility of super spreader events. In the interval $[t,t+ \d t)$ with conditional probability $\alpha \lambda^*_{1,i}$ one infection takes place and with conditional probability $\beta \lambda^*_{1,i}$ there are $k$ infections. The parameters $\alpha, \beta$ and $k$ are chosen such that the conditional mean is equal to $ \lambda^*_{1,i}$ and the conditional variance is equal to $ \gamma \lambda^*_{1,i}$. In our implementation of this bootstrap algorithm we have chosen $k$ equal to $\gamma +1$, with estimated variance factor $\gamma$ taking the value 65.3 for the infection daily data in France, and 141.4 for the weekly data described in the Supplementary Material.

When predicting hospitalizations, the bootstrap procedure needs to be slightly modified. In this case, the sample consists of pairs $\{ (N_{1,0},N_{2,0}), (N_{1,1},N_{2,1}), \ldots,(N_{1,T}, N_{2,T})\}$, where $N_{1,i}$ is the number of infections at the $i$th day, and  $N_{2,i}$ is the number of new hospitalized at the $i$th day, for $i=1,2,\ldots, T$. From these, we generate bootstrap samples including the following additional step in the algorithm above:
\begin{enumerate}
	\item [4.] Given $N_{1,0}, N^*_{1,1}, \ldots, N^*_{1,i-1}$, generate $N^*_{2,i} \longrightarrow Pois(\lambda^*_{2,i})$, with $\lambda^*_{2,i}=N_{1,0} \hat{\mu}_{2}(i,i)+ N^*_{1,1} \hat{\mu}_2(i,i-1)+\ldots +N^*_{1,i-1} \hat{\mu}_2(i,1)$.
\end{enumerate} 

}

\subsection{Forecasting with the  publicly reported reproduction numbers} \label{sec:fore_Rt}

Figure \ref{fig:foreInf_R} illustrates how publicly reported reproduction numbers from October 2020 can be used to forecast the number of infections during that month. { The results resemble those obtained using the {(infeasible) optimal $C_{t^*,h}$ value}, although they move slightly farther from the actual (adjusted) values in October. }
We follow the forecasting principle introduced in equation \eqref{eq:fore1} of Section \ref{sec:forecast}, but in this case we use the average of the reported  {$R_t$} numbers calculated with data not later than 31st of October, that is,  {$\bar R_{t^*}=1.34$}. More specifically, we predict number of infected every day in October 2020,  {taking $C_{t^*,h}=\bar R_{t^*}$, and  extrapolating the infection rate as $\widetilde{\widetilde\mu}_1\left((t^*+s)/T,u\right)=\widehat\mu_1\left(t^*/T,u\right) \times \left (1+ (\bar R_{t^*}-1)\frac{s}{h}\right )$} with $1\leq s \leq h$, and $h=31$.  
The graph highlights the potential of incorporating expert knowledge into our forecasting methodology.

{
\section{Comparison with existing models}\label{sec:comparison}

Several works in the recent literature have suggested either Hawkes models or SIR models  to analyse pandemic data (Bertozzi et al., 2020; Escobar, 2020, Garetto et al., 2021; Koyama et al., 2021; Kresin et al., 2022). Here we relate our approach to these works and argue that, while our approach makes use of Hawkes models, it is more flexible and general, allowing a more realistic description of the pandemic dynamics,  in particular at early stages of the pandemic when more detailed information is not available.

\cite{Bertozzi:et:al:20} considers a self-exciting counting process with intensity of the form $\lambda (t) = h(t) + \int R(t) g(t-t') N(\d t')$. 
In \cite{Koyama:et:al:21} $R(t)$ is replaced by $R(t-t')$. These papers propose estimates of $h(t)$ and $R(t)$ that heavily rely on parametric specifications of $g$. In \cite{Bertozzi:et:al:20} it is a Gamma or Weibull function. In \cite{Garetto:et:al21} $g$ is an indicator function of an interval. Formulation in \cite{Escobar:20} is similar but considering a decaying exponential with constant rate. In our model we replace $R(t) g(t-t')$ by $\mu(t/T, t-t')$ and thus, 
\begin{itemize}
\item [] (a) Our model is fully nonparametric.   This avoids biases of parametric models where the parametric assumptions may lead to inaccuracies which can be easily overseen and may lead to misinterpretations of the data.
\item [] (b) Identification is achieved by using different time scales of the first and second argument of $\mu$. Daily changes of $R$ are deferred to outside experts.
\item [](c) The estimation is derived by fitting nonparametrically the ratio $\# infected / \# exposed$ in an iterative procedure  where in each step the nominator or the denominator is updated. We compare this with estimators in survival analysis where the link between  $\# infected$ and $\# exposed$ is observed.
\end{itemize}

Besides avoiding biases, (a)--(c) increases the flexibility of the model. From a theoretical perspective, Hawkes processes often use the assumptions that : 
\begin{itemize}
\item [] (*) $h(t) \neq 0$
\item [] (**)  $\int R(t) g(t-t') \d t' < 1$.
\end{itemize}
Assumption (*) makes sense only at the start of an epidemic where in a region infections may be to a large extent ``imported''. In later stages of the epidemic one should assume $h(t) = 0$. Then the model is much better interpretable. The case $h(t) = 0$ is allowed in \cite{Garetto:et:al21}. If $h(t) = 0$  one needs during stable periods of the epidemic that  $\int R(t) g(t-t') \d t' > 1$, temporally.  Otherwise, the pandemic will exponentially fast turn down. Our approach  allows that $\int R(t) g(t-t') \d t'$ fluctuates around 1 which allow to have also stable periods of the pandemic  together with $h(t) =0$. 

\cite{Bertozzi:et:al:20} and \cite{Kresin:et:al:22} both explain the similarities between the well known SIR model and Hawkes processes. When comparing to noiseless SIR and SEIR models, these are solved by exponential functions, which restricts the models and their flexibility also for stochastic specifications of  SIR and SEIR models. 

It is important to remark that neither of these previous works address the missing link data problem that is so central for our  paper. For instance,  \cite{Bertozzi:et:al:20} models the time series of infected with the reproduction number directly in the formula, assuming indirectly a non-dynamic delay distribution. On the contrary, our approach provides an explicit model for the delay distribution that is more flexible and dynamic. The non-dynamic delay methodology of  \cite{Bertozzi:et:al:20} can also be found in the state-of-the-art works of \cite{Escobar:20}, \cite{Garetto:et:al21}  and \cite{Kresin:et:al:22}.

  We applied a bootstrap test for the null hypothesis of a non-dynamic delay distribution using a restricted version of our model comparable to the  approach of \cite{Bertozzi:et:al:20}. 
The hypothesis of non-dynamic delay  means that $ H_0:  \mu_1(t_1,u)=\mu_1(t_2,u)$, for all $0< u \leq t_1<t_2$. Thus under $ H_0$ the transfer function $\mu_1$ is constant over the notification date. Estimation and forecasting under $ H_0$ can be performed using a one-dimensional version of our local linear estimator in equation \eqref{eq:alpha1} as was done in \cite{Gamiz:etal:22}.  We make use of their estimator to test the hypothesis  $ H_0$.  The alternative allows that there is a calendar effect on the dynamics of infections, and thus, $\mu_1$ varies with  two time arguments: notification date ($t$)  and delay time ($0<u<t$). 
For this testing problem we used  the supremum norm between our estimator and the estimator of  \cite{Gamiz:etal:22}.  For calibration of the test we use a bootstrap procedure similar to the one described in the previous section but where now data were generated under the null hypothesis. Applying this testing procedure with 5000 bootstrap samples we obtain an empirical $p$-value of  0.0348. 
We performed a similar test for the hospitalization process. In this case we are interested in testing whether the function $\mu_2$ defined in \eqref{eq:haw2} is constant in its first argument, meaning that the dynamics of new hospitalizations does not change with notification date. Now we simulated bootstrap samples under this null hypothesis. Using $5000$ bootstrap samples we obtained an empirical \emph{p}-value of 0.0033, which provided strong evidence against the null hypothesis. Full details on both bootstrap tests can be found in Appendix \ref{app:test}. 

}


\section{Conclusion}\label{sec:conc} 

This paper has developed a new time dependent and therefore dynamic principle where the entering observations are allowed to change definitions over time as long as this change is happening in a smooth manner. This allows us to work on otherwise problematic infection and hospitalization data that were omnipresent in the recent Covid-19 pandemic. One important feature of our modelling technique is that it only relies on relatively simple available data and therefore could serve as the first model used in the beginning of a pandemic. 
The data definition would be simple to communicate across borders and across scientific communities and our new approach could therefore serve as a useful benchmark method when the state of the pandemic is new and confusing. 
While our method is applicable to other statistical modelling problems than pandemic, we will below quickly mention the relationship between our work and some of the key academic works on modelling the pandemic. 
We consider \cite{Quick:etal:21} to be a state-of-the-art paper published during the pandemic. They work with the development of infections and do have very interesting methodology to estimate the $R$-number that we also work with in this paper. 
\cite{Quick:etal:21} also work with simple available data, but in contradiction to our work, they build a sophisticated model of the underlying mechanism. 
See also \cite{Jewell:21} and \cite{Dean:Yang:21} for excellent discussions and intuitive explanation of \cite{Quick:etal:21}. 
However, such a detailed micro-model does have the consequence of being limited to the problem at hand. 
We consider our approach to be simpler, with less bias and well fit for immediate implementation in the beginning of a pandemic. 
We have left the calculation of the $R$-number for external experts that might have prior knowledge worth incorporation. It is of course also fully aligned with our approach to fit the $R$-number via a regression methodology as suggested in \cite{Quick:etal:21}. 
\cite{Mukherjee:22} and \cite{Lin:22} suggest that communication and general cooperation between scientific fields is a key ingredient that should be developed early on in a pandemic. 
The communication of our method is easy, because both input and output are easy to understand, and our method is not depending on underlying assumptions that are difficult to communicate. On the other hand, we isolate the scientific cooperation in the single number used in the forecasting (the $R$-number) such that experts on the $R$-number do not need to be experts in the general scientific modelling of the problem.

We have developed the R package \emph{pandemics}, which is already available on CRAN  \citep{Gamiz:etal:24} and implements the methodology in both papers.


	\appendix
	\section{Asymptotic analysis}\label{sec:theo1} 
	In this section we come back to 
	our setting of Section \ref
	{sec:mod1} where the estimator $ \widehat \mu_1 (t/T, t-s)$ was defined as solution 
	of \eqref{estdefeq}. In this section we will argue that $ \widehat \mu_1 (t/T, t-s)$ is a consistent  estimator. This will be done in an asymptotic framework where the factor $n$ converges together with $T$ to infinity. We restrict ourselves here to a purely heuristic discussion. The Hawkes model of this paper is rather complex from a mathematical point of view. Typically one assumes in Hawkes processes that $\rho_1(t) > 0$ for the whole interval $t \in [0,T]$ and that $\int \mu_1(t/T, u) \d u < 1$. Then in average less than one person is infected by one infected person and the epidemic would stop if not infected persons with an incidence rate $\rho_1(t) $ would enter from outside. Typically, this is not an adequate model for an epidemic and it is more appropriate to assume that $\rho_1(t) = 0$. Then to keep the number of infections stable one needs that $\int \mu_1(t/T, u) \d u$ is greater equal 1 for some values of $t$. This complicates the mathematics. A rigorous mathematical analysis of our model under this assumption is out of the scope of this paper. We only will present a heuristic discussion why it is expected that $ \widehat \mu_1$ is a consistent estimator of $  \mu_1$. We will argue that, approximately, $ \widehat \mu_1$ is the solution of a linear integral equation, see \eqref{eq:add4}, \eqref{eq:add5}. In particular, in our heuristic study we will not discuss under which conditions this integral equation has a unique solution. 	
	
	We will see that $n$ is the order of the number of infections in a finite interval, under the assumptions that we specify now.
	For stating the assumptions we will make use of the martingale $M_1$ defined by $M_1(\d t) = N_1(\d t) - \lambda_1(t) \d t$, 
	see  \cite{Andersen:etal:93}.
		
	Furthermore, put $\chi_1(t/T,t-s) = \sum_{k=1} ^{\infty} \mu_1^{(*k)}(t/T, t-s) $,
	with $\mu_1^{(*k)}(t/T, t-s) = \int_0^t \mu_1(t/T, t-v) \mu_1^{(*(k-1))}(v/T,v-s) \d v$, and $\mu_1^{(*1)}(t/T, t-s)=\mu_1(t/T, t-s)$.  Note that $\mu_1^{(*2)}(t/T, t-s) \mathrm d t$ is the probability that a person infected at $s$ infects a person that infects a further person in $(t, t+ \mathrm d t]$. Similarly, $\mu_1^{(*k)}(t/T, t-s) \mathrm d t$ is defined as probability with $k-1$ intermediate infections. Finally, $\chi_1(t/T,t-s) \mathrm d t$ is the probability that a person infected at $s$ starts a series of infections which leads to an infection of  a person  in $(t, t+ \mathrm d t]$.
		
	By iterative application of the Hawkes model equation one can show that 
	\begin{eqnarray} \label{eq:krlstor1}
		\Lambda _1(t) &=& \E [\lambda_1(t)] = n \rho_1(t) + n  \int_0 ^t \chi_1(t/T,t-u) \rho_1(u) \d u, \\
		N_1(\d t) &=& M_1(\d t) + \d t \ \bigg ( \int_0 ^{t^-} \chi_1(t/T,t-u)M_1( \d u) +  n \rho_1(t)\nonumber  \\
		&& \qquad+ n \int_0 ^t \chi_1(t/T,t-u) \rho_1(u) \d u\bigg ),\label{eq:krlstor2}\\
		\E [N_1(\d t)] &=& \mathrm d t
		  \bigg ( n \rho_1(t) + n \int_0 ^t \chi_1(t/T,t-u) \rho_1(u) \d u\bigg ), \label{eq:krlstor3}\\
		  \E [N_1(\d t)N_1(\d t^\prime)] &=&  \mathrm d t  \mathrm d t^\prime \bigg ( \Lambda _1(t) \Lambda _1(t^\prime) + \chi_1(t/T,t-t^\prime) \nonumber \\
		  && + \int _0 ^{t^\prime}  \chi_1(t/T,t-s) \Lambda _1(s)  \chi_1(t^\prime/T,t^\prime-s) \mathrm d s\bigg ) \nonumber\\
		  && \text{ for } t^\prime < t, \label{eq:krlstor4}\\
		   \E [N_1(\d t)N_1(\d t)] &=& \Lambda _1(t)  \mathrm d t.
 \label{eq:krlstor5}
	\end{eqnarray}
	
	 For a proof and discussion of these equations see \cite{Mammen:Muller:23}. We assume that the term $ \chi_1(t/T) =\int_0 ^t \chi_1(t/T,t-u) \rho_1(u) \d u + \rho_1(t)$ is bounded away from 0 and from infinity on the interval $[0,T]$. In particular, this is equivalent to assume  that the expected number of infections in a finite interval is of order $n$. For the martingale we have that $E[M(t+\delta) - M(t)]^2= O(\delta n )$. To see this one can  construct Hawkes processes by a modification of the so-called cluster process. Generate for $j=1,...,n$ independent Poisson processes with intensity {$\rho_1(t)$}. Interpret the arrivals of the Poisson processes as incoming infected persons and denote by $N_{1,j}(t)$ the  number of persons infected by these persons or infected  by them  indirectly  over a finite line of infections. This gives independent processes. Note that $N_{1}(t)$ can be understood as $\sum_{j=1} ^n N_{1,j}(t)$. 
	
	We consider the estimator $\widehat \mu_1(t/T, t-s)$ at  points $t = T x$, for  fixed choices $0 < x < 1$. 	
	Furthermore, we assume that $\widehat \mu_1$ is chosen such that it is uniformly Lipschitz continuous with respect to its first and second argument. One can see that there exist solutions of  \eqref{estdefeq} with this property. One theoretical example is the theoretical choice $\widehat \mu_1= \mu_1$ as can be seen by some careful lengthy considerations. Thus we are not speaking about an empty set of estimators.
	
	From \eqref{estdefeq} we have that 
	\begin{eqnarray} \label{eq:add0}&& \int_{0 \leq v < u \leq T} {\rm C}_1(s,t,v,u) K_{1,b_1} \left ( \frac {t-u} T\right ) K_{2,b_2} (t-s - (u-v))
		\\ && \qquad \times \frac {\widehat \mu_1 (u/T, u-v)} {\widehat \mu_1 (t/T, t-s)} \frac { N_1(\d v) N_1(\d u)}
		{\int_0^{u^-}\widehat \mu_1(u/T, u-w)  N_1(\d w)}  \nonumber  \\ && \qquad \times \frac 1 {  \int_{0< u \leq T} 
			\bar w_{s,t}(u) 
			\d u} = 1 +o_P(1), \nonumber
	\end{eqnarray}
	where $\bar w_{s,t}(u) = \int_{0\leq v < u} {\rm C}_1(s,t,v,u) K_{1,b_1} \left ( \frac {t-u} T\right ) K_{2,b_2} (t-s - (u-v))
	N_1(\d v) $.
	
	In a first step we will show the following claim: 
	\begin{eqnarray} \label{eq:add1}
	\int_0^{u^-}\widehat \mu_1(u/T, u-w)  \frac 1 n N_1(\d w)  = \int_0^{u^-}\mu_1(u/T, u-w)  \frac 1 n N_1(\d w)+ o_P(1).
	\end{eqnarray}
	We will show \eqref{eq:add1} under the additional assumption that
	 $K_1$ and $K_2$ have a bounded support, say $[-1,1]$. Then using Lipschitz continuity of $\widehat \mu_1$ we get for $|t-u| \leq b_1 T$, $|t-s - (u-v)| \leq b_2$ that \begin{eqnarray} \label{eq:add2} \widehat \mu_1 (u/T, u-v) =  \widehat \mu_1 (t/T,t-s)+ o_P(1).\end{eqnarray} This can be used to show that
	that $\widehat \mu_1$  solves 
	\begin{eqnarray*} && \int_{0 \leq v < u \leq T} {\rm C}_1(s,t,v,u) K_{1,b_1} \left ( \frac {t-u} T\right ) K_{2,b_2} (t-s - (u-v))
		\\ && \qquad \times \frac { N_1(\d v) N_1(\d u)}
		{\int_0^{u^-}\widehat \mu_1(u/T, u-w)  N_1(\d w)}   \\ && \qquad \times \frac 1 {  \int_{0< u \leq T} 
			\bar w_{s,t}(u) 
			\d u} = 1 +o_P(1). \nonumber
	\end{eqnarray*}
	 We now use that $N_1({\rm d}  u)= (\int_0^{u^-} \mu_1(u/T, u-w) n^{-1} N_1({\rm d} w)) {\rm d} u+  M_1({\rm d}  u)$. This gives by some calculations
	\begin{eqnarray*} && \int_{0 \leq  u \leq T}w_{s,t}(u) \frac {\int_0^{u^-} \mu_1(u/T, u-w)  n^{-1}  n^{-1} N_1(\d w)}
		{\int_0^{u^-}\widehat \mu_1(u/T, u-w)  N_1(\d w)}  	 \d u= 1 +o_P(1), \nonumber
	\end{eqnarray*}		
	where $w_{s,t}(u) = \bar w_{s,t}(u)  /  \int_{0 \leq  r \leq T} 	\bar w_{s,t}(r)  \d r$.	With the help of   \eqref{eq:krlstor1}-- \eqref{eq:krlstor5} one can show that  \begin{eqnarray} \label{eq:add3}
w_{s,t}(u) = K_{1,b_1} \left ( \frac {t-u} T\right ) \left (1 +o_P(1) \right ). \end{eqnarray} This can be used to show \eqref{eq:add1}.
	
	We now use  \eqref{eq:add0}--\eqref{eq:add3} to show the following linear integral equations for $\widehat \mu_1$ with random kernels. 
\begin{eqnarray*} && \widehat \mu_1 (t/T, t-s) - \int_{0 \leq v < u \leq T} {\rm C}_1(s,t,v,u) K_{1,b_1} \left ( \frac {t-u} T\right ) K_{2,b_2} (t-s - (u-v))
		\\ && \qquad \times  {\widehat \mu_1 (u/T, u-v)} {} \frac { N_1(\d v) N_1(\d u)}
		{\int_0^{u^-} \mu_1(u/T, u-w)  N_1(\d w)}  \nonumber  \\ && \qquad \times \frac 1 {  \int_{0< u \leq T} 
			\bar w_{s,t}(u) 
			\d u} = o_P(1), \\
	&&\int_0^{u^-}\widehat \mu_1(u/T, u-w)  \frac 1 n N_1(\d w)  = \int_0^{u^-}\mu_1(u/T, u-w)  \frac 1 n N_1(\d w)+ o_P(1).
	\end{eqnarray*}
	Using again \eqref{eq:add1}--\eqref{eq:add3} one can argue that the random kernels can be replaced by deterministic functions:
	\begin{eqnarray} \nonumber&& \widehat \mu_1 (t/T, t-s) - \int_{0 \leq v < u \leq T} {\rm C}_1(s,t,v,u) K_{1,b_1} \left ( \frac {t-u} T\right ) K_{2,b_2} (t-s - (u-v))
		\\ \nonumber && \qquad \times  {\widehat \mu_1 (u/T, u-v)} {} \frac { \E[N_1(\d v) N_1(\d u)]}
		{\int_0^{u^-} \mu_1(u/T, u-w)  \E[N_1(\d w)]}  \nonumber  \\ && \qquad \times \frac 1 {  \int_{0< u \leq T} 
			 \E[\bar w_{s,t}(u) ]
			\d u} = o_P(1), \label{eq:add4} \\
	&&\int_0^{u^-}\widehat \mu_1(u/T, u-w)  \frac 1 n  \E[N_1](\d w)  = \int_0^{u^-}\mu_1(u/T, u-w)  \frac 1 n  \E[N_1](\d w) \nonumber\\
	&& \label{eq:add5} \qquad + o_P(1).
	\end{eqnarray}
	We conjecture that under assumptions on the deterministic kernel of this integral equation one can conclude that $ \widehat \mu_1 =   \mu_1 + o_P(1).$ We expect that this line of arguments can be made rigorous but this would require a very detailed and complex mathematical analysis.

	For the asymptotic study of our estimator it is helpful to assume that $N_1$ is a Hawkes process. But essentially we only use that $\E [N_1(\d t)\mid {\mathcal F}_1(t)]=\left(\int_0^{t^-} \mu_1(t/T,t-s)N_1(\d s)\right)\d t$ + $n \rho_1(t)\d t$, 
	where not necessarily $N_1(\d t)$ must be $\leq 1$ almost surely. In the case of epidemic data one may wish to allow for super spreader events which would mean that $N_1(\d t) \gg 1$. In this case one has to deviate from a Hawkes model. On the other side for large  values of $n$ these differences may not be significant after discretization of the data. In  Section \ref{sec:mod2} we also already deviated from a Hawkes model if we use $N_1$ to model  exposure for another process $N_2$.
	
	\section{Derivation of the local linear estimator with full information} \label{ap:estim}

	With full information, we observe the processes $ N_{1,1}(t,\d s), \ldots, N_{1,n}(t,\d s)$ for $ s < t \leq T $, corresponding to $n$ subjects that are followed over period of time $(0,T]$. For $i =1,\ldots, n$, $N_{1,i}( t,\d s)$ counts the number of individuals infected by person $i$ in the interval $(s, t]$, given that he/she was infected in $(s, s+ \d s)$, with $0<s<t \leq T$.  
	
	For $i=1,\ldots, n$, let us define $Z_{1,i}(s)$ an indicator variable, such that $Z_{1,i}(\d s)$ jumps 1 if person $i$ becomes infected in $(s, s+\d s)$. 
	Additionally, we denote $N_{1,i}(\Delta u,\cdot)=N_{1,i}(u+\Delta,\cdot)-N_{1,i}(u,\cdot)$, for $\Delta >0$.

The local linear estimator of the infection rate $\mu_1$, defined in Section \ref{sec:mod1}, is obtained minimizing the function 
\begin{eqnarray}\label{eq:estim}
\nonumber&&\hspace{-1.5cm} \sum_{i=1}^n\!\int_0^T\!\!\!\int_v^T\!\!\left( N_{1,i}( \Delta  u, \d v)- \theta_0 - \theta_{11}\left(\frac{t-u}{T}\right)-\theta_{12}(t-s-(u-v))\right)^2 \times  \\
&&\hspace{1.5cm}  K_{1,b_1} \left ( \frac {t-u} T\right ) K_{2,b_2} (t-s - (u-v)) \d u Z_{1,i}( \d v),
\end{eqnarray}
with $K_{j,b_j}(\cdot)=K_j(\cdot/b_j)/b_j$, where $K_j$ is a one-dimensional kernel function, $j=1,2$. Here we use the notation  $\int N_{1,i}(\Delta u,\cdot)g(u)\d u= \int g(u)N_{1,i}(\d u,\cdot)$, for any function $g$. The above optimization problem leads to the following equations:
\begin{eqnarray}
\nonumber	&&\theta_0{\rm a}_0(t,s)+\theta_{11}{\rm a}_1(t,s) \ + \  \theta_{12}{\rm a}_2(t,s) \ - \ {\rm b}_0(t,s) =0\\
	&&\theta_0{\rm a}_1(t,s)+\theta_{11}{\rm A}_{11}(t,s)+\theta_{12}{\rm A}_{12}(t,s)-{\rm b}_1(t,s)=0\\
\nonumber	&&\theta_0{\rm a}_2(t,s)+\theta_{11}{\rm A}_{12}(t,s)+\theta_{12}{\rm A}_{22}(t,s)-{\rm b}_2(t,s)=0
\end{eqnarray}
where
\begin{eqnarray}\label{eq:a_i}
	\nonumber && \hspace{-2cm}{\rm a}_{l}(t,s)=\int_{0\leq v < u \leq T}\! K_{1,b_1}\left(\frac{t-u}{T}\right) K_{2,b_2}(t-s-(u-v))
	\\ && \qquad   \qquad \times \left(\frac{t-u}{T}\right)^{\delta_{l1}}(t-s-(u-v))^{\delta_{l2}} N_{1}(\d v) \ \d u,
\end{eqnarray}
and 
\begin{eqnarray}\label{eq:b_i}
	\nonumber && \hspace{-2cm}{\rm b}_{l}(t,s)=\int_{0\leq v < u \leq T}\! K_{1,b_1}\left(\frac{t-u}{T}\right) K_{2,b_2}(t-s-(u-v))
	\\ && \qquad   \qquad \times \left(\frac{t-u}{T}\right)^{\delta_{l1}}(t-s-(u-v))^{\delta_{l2}} N_{1}(\d u,\d v),
\end{eqnarray}
for $l=0,1,2$, with $\delta_{lh}=1$ if $l=h$, and 0, otherwise; and
\begin{eqnarray}\label{eq:A_ij}
	\nonumber &&\hspace{-2cm} {\rm A}_{l,m}(t,s)=\int_{0\leq v < u \leq T}\! K_{1,b_1}\left(\frac{t-u}{T}\right)K_{2,b_2}(t-s-(u-v))
	\\ &&  \quad \times \left(\frac{t-u}{T}\right)^{\delta_{l1}+\delta_{m1}}(t-s-(u-v))^{\delta_{l2}+\delta_{m2}} N_{1}(\d v) \ \d u,
\end{eqnarray}
for $l,m=1,2$, with $\delta_{lh}$ defined above, and $\delta_{mh}=1$ if $m=h$ and 0 otherwise.

We have used that $\sum_{i=1}^nN_{1,i}(\d u, \d v) Z_{1,i}(\d v)= N_{1}(\d u, \d v)$, that is, we obtain the total number of persons infected in $(u, u+ \d u)$ by any person who was infected in $(v, v+ \d v)$ (see notation in Section \ref{sec:mod1}). Also, we have that $\sum_{i=1}^nZ_{1,i}(\d v)=N_1(\d v)$ counts the number new infections in $(v, v+ \d v)$. 

Direct solution of the above equations in $\theta_0$ gives the following expression for the local linear estimator of $\mu_1$:
\[
\widehat{\mu}_1(t,s)= \frac{{\rm b}_0(t,s)-{\rm \bf b}(t,s)^t{\rm \bf A}(t,s)^{-1}{\rm \bf a}(t,s)}
	{{\rm a}_0(t,s)-{\rm \bf a}(t,s)^t{\bf {\rm \bf A}}(t,s)^{-1}{\rm \bf a}(t,s)} :=\frac{O_1(t,s)}{E_1(t,s)},
\]
where ${\rm \bf b}$ is a vector function with components ${\rm b}_l$, and  ${\rm \bf a}$ is a vector function with components ${\rm a}_l$, with $l=1,2$, and  ${\rm \bf A}$ is a matrix function with entries ${\rm A}_{lm}$, with $l,m=1,2$.
Some easy calculations lead to the expression of the estimator as an intuitive ratio of the smoothed occurrences and smoothed exposures  as it is expressed in equation \eqref{eq:alpha1} given in Section \ref{sec:mod1}. In particular it is not difficult to check that 
\[
O_1(t,s)=
\int_{0\leq v < u \leq T}{\rm C}_1(s,t,v,u) \ K_{1,b_1}\left(\frac{t-u}{T}\right) K_{2,b_2}(t-s-(u-v)) N_{1}(\d u, \d v),
\]
and, 
\[
E_1(t,s)=
\int_{0\leq v < u \leq T}{\rm C}_1(s,t,v,u) \ K_{1,b_1}\left(\frac{t-u}{T}\right) K_{2,b_2}(t-s-(u-v)) N_{1}(\d v) \ \d u
\]
with ${\rm C}_1(s,t,v,u)=1-\big ((t-u)/T,t-s-(u-v)\big ) {\rm {\bf A}}(t,s)^{-1}{\rm {\bf a}}(t,s)$.\\

Finally, the estimation of the hospitalization rate $\mu_2$ defined in Section \ref{sec:mod2}, with full information, is based on the observation of the processes $ N_{2,1}(t,\d s), \ldots, N_{2,n}(t,\d s)$ for $ s < t \leq T $,  from $n$ subjects followed over period of time $(0,T]$. For $i =1,\ldots, n$, $N_{2,i}( t,\d s)$ takes value 1 if individual $i$, infected in  $(s, s+ \d s)$, is hospitalized in the interval $(s,t]$, and 0 otherwise, with $0<s<t \leq T$. Then the local linear estimator is obtained minimizing the following function: 

\begin{eqnarray}\label{eq:estim}
	\nonumber&&\hspace{-1.5cm} \sum_{i=1}^n\!\int_0^T\!\!\!\int_v^T\!\!\left( N_{2,i}( \Delta  u, \d v)- \theta_0 - \theta_{11}\left(\frac{t-u}{T}\right)-\theta_{12}(t-s-(u-v))\right)^2 \times \\
	&&\hspace{1.5cm}  K_{1,b_1} \left ( \frac {t-u} T\right ) K_{2,b_2} (t-s - (u-v))  \d u Z_{1,i}( \d v).
\end{eqnarray}
{ 
\section{Testing for non-dynamic delay distributions of infections and hospitalizations}  \label{app:test}

Let  $N_{1,0}, N_{1,1}, \ldots, N_{1,T}$ be daily observations of the process $\{N_1(t); t>0\}$, where $N_{1,i} $ is the number of infections on the $i$th day ($i=1,2,\ldots, T$). Let assume that the intensity function of the process follows a Hawkes model with intensity function given in \eqref{eq:haw1} where the baseline intensity is assumed $\rho_1=0$. 
We are interested in solving the following testing problem: 
\begin{eqnarray*}
	&&H_0:   \mu_1(s,u)=\mu_1(t,u) , {\rm for \ all } \ 0< u \leq s<t; \\
	&&H_1:  \mu_1(s,u)\neq\mu_1(t,u), {\rm for \ some } \ s<t, \ {\rm and } \ 0<u\leq s.
\end{eqnarray*}
Under the null hypothesis, $\mu_1$ is constant over the first argument (notification date, $t$) and therefore the delay distribution is stationary. On the contrary, under the alternative, $\mu_1$ varies with the two time arguments: notification date ($t$)  and delay time ($0<u<t$).

To solve the above testing problem we propose the following bootstrap procedure.
\begin{enumerate}
	\item[1.] From the original data, $N_{1,0},\ldots, N_{1,T}$, estimate  $\mu_1$ under the null hypothesis and obtain $\widehat{\mu}_{1,H_0}(t_i)$,  for $t_i = i/T$ with $i\in \{1,\ldots, T\}$.
	\item[2.]  From the original data, $N_{1,0},\ldots, N_{1,T}$, estimate  $\mu_1$ under the the alternative  hypothesis and obtain $\widehat{\mu}_{1,H_1}(t_j,t_i)$, for $i,j \in \{1,\ldots, T\}$, and $i <j$.
	\item [3.] Compute the test statistics as the supremum-norm distance between  $\widehat{\mu}_{1,H_0}$ and $\widehat{\mu}_{1,H_1}$ as
	\begin{equation} \label{eq:D}
		D_{\infty}= \underset{1\leq i \leq j \leq T}{\max}\left\{\vert \widehat{\mu}_{1,H_0}(t_i)-\widehat{\mu}_{1,H_1}(t_j,t_i)\vert\right\} .
	\end{equation}
	\item [4.] Generate a bootstrap sample under the null hypothesis. That is, given $N^*_{1,0},N_{1,0}, N^*_{1,1}, \ldots, N^*_{1,i-1}$, simulate $N^*_{1,i} \rightarrow Pois(\lambda^*_{1,i})$, with $\lambda^*_{1,i}=N_{1,0}^* \hat{\mu}_{1,H_0}(t_i)+ N^*_{1,1} \hat{\mu}_{1,H_0}(t_{i-1})+\ldots +N^*_{1,i-1} \hat{\mu}_{1,H_0}(t_1)$
	\item [5.]  Based on the bootstrap sample $N^*_{1,0}, ...,N^*_{1,T}$, compute a bootstrap version of the estimator $\widehat{\mu}^*_{1,H_0}$, under the null hypothesis, and $\widehat{\mu}^*_{1,H_1}$, under the alternative. 
	\item [6.] Derive a bootstrap version of the test statistics, $D^*_{\infty}$, as \eqref{eq:D} but from the bootstrap sample.
	\item [7.] Repeat steps 4-6 a large number of times $B$ and approximate the distribution of $D_{\infty}$ under the null hypothesis by Monte Carlo from the bootstrap replicates $D^*_{\infty,1},\ldots, D^*_{\infty,B}$.

\end{enumerate}

The test above can be adapted also for the hospitalization process. In this case we are interested in testing whether the function $\mu_2$ defined in \eqref{eq:haw2} is constant in its first argument, meaning that the dynamic of new hospitalizations does not change with notification date. 
In this case the sample is made of pairs $\{ (N_{1,0},N_{2,0}), (N_{1,1},N_{2,1}), \ldots,(N_{1,T}, N_{2,T})\}$, where $N_{1,i}$ is the number of infections at the $i$th day; and,  $N_{2,i}$ is the number of new hospitalized at the $i$th day, for $i=1,2,\ldots, T$. We formulate the following testing problem: 
\begin{eqnarray*}
	&&H_0:   \mu_2(t_1,u)=\mu_2(t_2,u) , {\rm for \ all } \ 0< u \leq t_1<t_2; \\
	&&H_1:  \mu_2(t_1,u)\neq\mu_2(t_2,u), {\rm for \ some } \ t_1<t_2, \ {\rm and } \ 0<u<t_1.
\end{eqnarray*}
To solve this problem we use a similar bootstrap procedure as for the infection process above, but simulating bootstrap samples of hospitalized conditioned on the infections. More specifically, we replace step 4 above, by the following:

\begin{enumerate}
	\item[4'.] Generate bootstrap samples of new hospitalized under the null hypothesis conditioned on the registered number of daily infected. That is, given $N_{1,0}, N_{1,1}, \ldots, N_{1,i-1}$, simulate $N^*_{2,i} \rightarrow Pois(\lambda^*_{2,i})$, with $\lambda^*_{2,i}=N_{1,0} \hat{\mu}_{2,H_0}(t_i)+ N_{1,1} \hat{\mu}_{2,H_0}(t_{i-1})+\ldots +N_{1,i-1} \hat{\mu}_{2,H_0}(t_1)$.
\end{enumerate}

}

{
\section{Data adjustment for variations by day of the week} \label{app:koyama}

The analysis of Covid-19 pandemic data presented in this paper reveals consistent variations in reported cases throughout the week, with notably fewer cases typically recorded on weekends. To correct for this effect we apply a weekday adjustment similar to the method suggested in \cite{Koyama:et:al:21}. Given a sequence of daily case counts $\{N_i;i=1,\ldots,T\}$, where $T$ is the total number of days, we define a mapping from each day $i \in \{1,\ldots, T\}$ to a weekday index, i.e. $ d(i) \in \{1, \ldots, 7\}$, with Monday as 1 and Sunday as 7 (or any consistent 7-day indexing), so that, for example, if day $i$ falls on a Monday, then $d(i) = 1$, and so on.

For each day of the week, we compute the total number of reported cases

\[
f_d = \sum_{\substack{i = 1 \\ d(i) = d}}^T N_i,\quad \text{for } \  d \in \{1, \ldots, 7\}.
\]
Let $n = \sum_{i=1}^T N_i$ denote the total number of reported cases across all days. The weight for day $i$ is then defined as

\[
w_{d(i)} = \frac{f_{d(i)}}{n/7}, \quad \text{for }   i=1,\ldots,T.
\]

These weights represent the relative contribution of each weekday to the total count, scaled such that their average is 1, that is $(1/7)\sum_{d} w_d=1$. If no weekday effect is present, all weights $w_d$ would be close to 1.

Finally, the adjusted daily counts $\{\tilde{N}_i, i=1,\ldots, T\}$ are obtained by dividing each original count by its corresponding weekday weight

\[
\tilde{N}_i = {N_i}/{w_{d(i)}} \quad \text{for } i = 1, \ldots, T.
\]

This procedure reduces the influence of systematic underreporting or overreporting on specific weekdays, making the adjusted series more suitable for time-series analysis and modelling.

}

\section{Simulations} \label{sec:simu}

We present a brief simulation study supporting our estimation and forecasting methods. Following the model description in Section \ref{sec:mod1}, we simulate the process $N_1(t)$, consisting of the number of persons infected in the interval $(0,t]$, in a time grid of $t=1,..., T=100$ days. We assume that the process intensity is of the form in equation \eqref{eq:haw1}, and, for convenience, we write here the excitation function as $\mu_1(t,s)=\mu_0(t/T,(t-s+1)/T)/T$, for $1\leq s \leq t \leq 100$. With this formulation, the second argument of $\mu_1$ does not refer to duration time, and duration time is given by $t-s+1$. Additionally,  we assume that the support for duration time is $D_1=T$. 

\begin{figure}[htp]
	\begin{center}
	\includegraphics[width=0.49\textwidth]{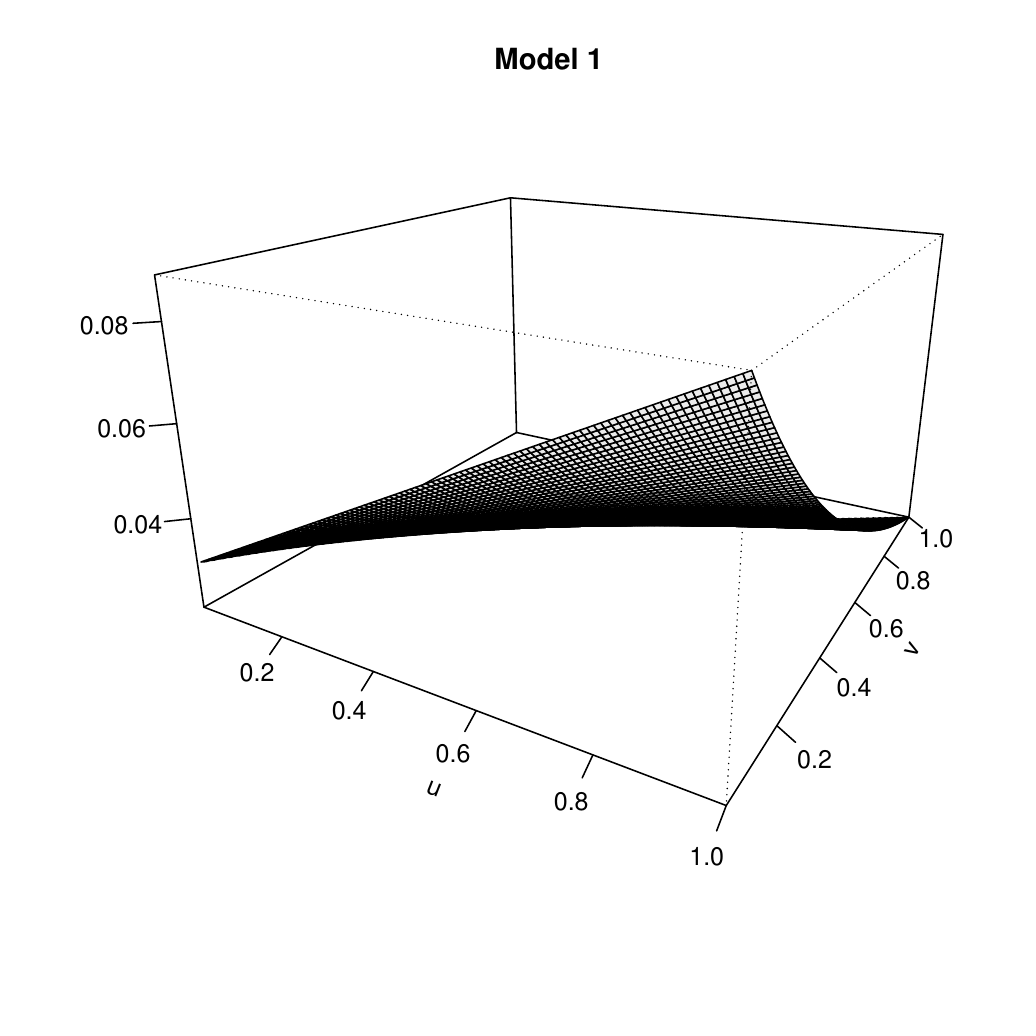}
		\includegraphics[width=0.49\textwidth]{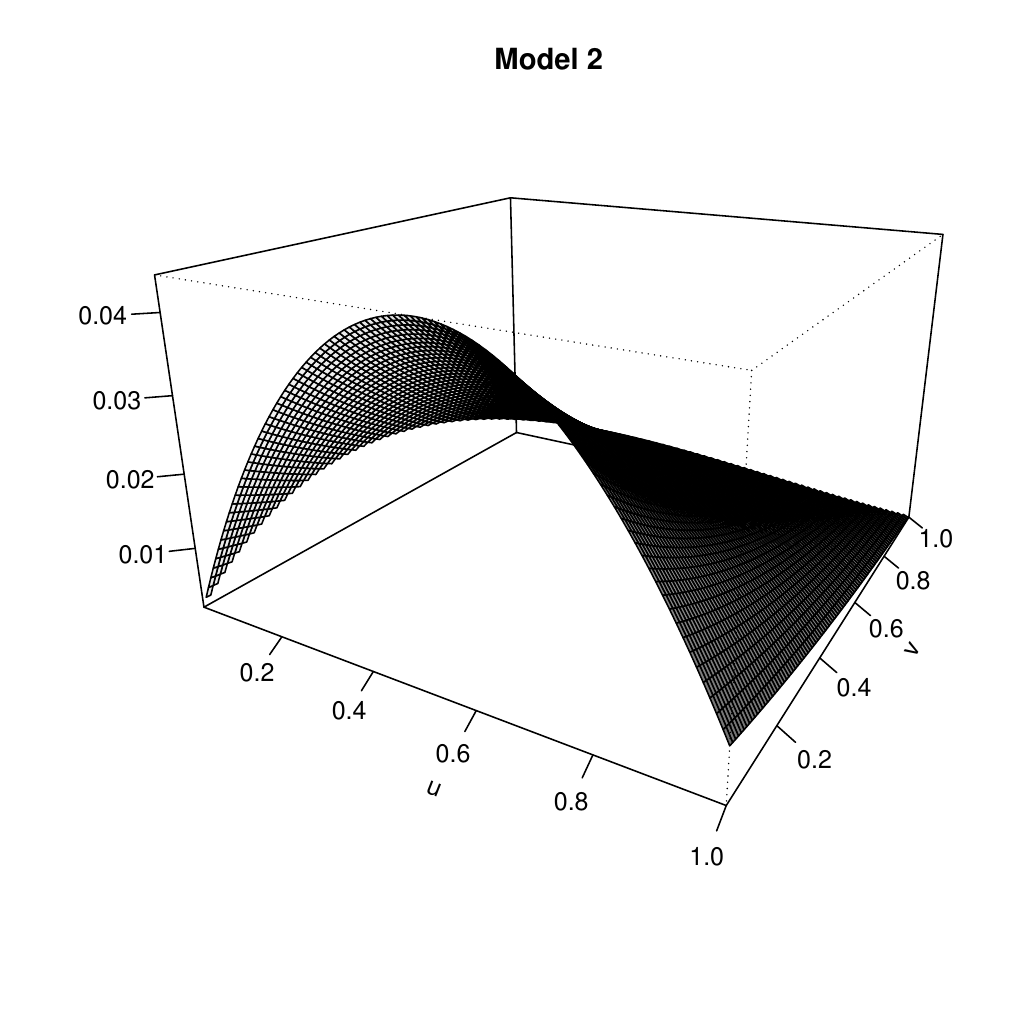}
	\caption{True models for infections in the simulation study.}\label{fig1}
	\end{center}
\end{figure}

We start the process  $N_1(t)$ with $n_0=1000, 10000, 100000$ infected persons. For the excitation function, $\mu_0$, we consider two models:
\begin{eqnarray*}
	{\rm Model \ 1:}&&\hspace{-.5cm}\mu_{0}(u,v)=(1+2 u)(3 e^{-1.5 v});\\ 
{\rm Model \ 2:}&&\hspace{-.5cm}\mu_{0}(u,v)=5.7 u(1-0.95u)(3 e^{-1.5v}),
\end{eqnarray*}
for $0<v<u\leq 1$, and 0 otherwise.  Figure \ref{fig1} shows graphs of the two models.

We have simulated 100 samples from each model and sample size. From each sample, we have computed two estimators of $\mu_1$, 
one based on full information and the other one based on partial information, as described in Section \ref{sec:estim} . Using these estimators and the forecasting principles described in Section \ref{sec:forecast}, we have computed prediction on days $101,...,105$. To evaluate the accuracy of predictions, we compare with the simulated data on the days  $101,\ldots, 105$, which were generated assuming that $\mu_1(t,s)=\mu_1(100,s)$, for all $t=101,\ldots, 105$, and  $s=1,\ldots, 100$, and 0 otherwise. This means assuming that the future behaves exactly as the immediate past. \\

Table  \ref{tab:pe} shows the prediction error, defined as $\sum_{t=101}^{105} (\widehat N_1(t) - N_1(t))^2 / \sum_{t=101}^{105} N_1(t)^2 $, where $\widehat N_1(t) $ is the predicted value of $N_1(t)$ based on partial or full information, respectively.
The bandwidths for the estimators were chosen by minimizing the prediction error. 
\begin{table}[htp]
	\centering
	\caption{Simulation results based on 100 simulations. {Numbers in the table consist of average  prediction errors  for both models (multiplied by $10^4$ for convenience).}}
	\begin{tabular}{ccc}
		\hline
		\multicolumn{3}{c}{Model 1 }\\
		$n_0$	& Partial Information & Full Information \\ 
		\hline
		1000 &1.3084 & 1.2289 \\
	    10000& 0.1861& 0.1082 	 \\    
        100000 & 0.09224& 0.0112 \\
		\hline
		\multicolumn{3}{c}{Model 2 }\\
		$n_0$	& Partial Information & Full Information \\ 
		\hline
		1000 &  207.0112 &205.2344 \\
		10000 &25.9729 &23.2703 \\
		100000 &5.3960 &2.2582    \\
		\hline
	\end{tabular}\label{tab:pe}
\end{table}

Table \ref{tab1} shows the average squared error of the estimators of $\mu_1$ given as $
ISE(\widehat{\mu}_{1,n_0})=n_0^{-1}\sum_{t=1}^{100}\sum_{s=1}^{t} (\widehat{\mu}_{1,n_0}(t,s)-\mu_1(t,s))^2$. As evidenced by the results shown in the tables, our approach yields accurate forecasts and low estimation errors, validating the empirical robustness of our method. Estimating with partial information leads to a loss in terms of prediction and estimation error, compared to the case of full information, as expected.

\begin{table}[H]
	\centering
\caption{Simulation results based on 100 simulations. For each model and sample size the averaged ISE values (MISE) are presented. Numbers have been multiplied by $10^5$ for convenience.}
	\begin{tabular}{ccc}
		\hline
		\multicolumn{3}{c}{Model 1 }\\
	$n_0$	& Partial Information & Full Information \\ 
		\hline
	1000 & 191.0628 & 2.9246 \\
	10000 & 19.0916 &  0.0432 \\ 
    100000 & 1.9089& 0.0007 \\

		\hline
		\multicolumn{3}{c}{Model 2 }\\
		$n_0$	& Partial Information & Full Information \\ 
		\hline
		1000 & 17.5122 &3.2323 \\
		10000 &1.7362 & 0.1786\\
        100000 &0.1713 &0.0103\\	
		\hline
	\end{tabular}\label{tab1}
\end{table}



\begin{thebibliography}{10}

\bibitem[Andersen et~al.(1993)]{Andersen:etal:93} {Andersen, P., Borgan, O., Gill, R. and Keiding, N.} (1993).
\newblock {\em Statistical Models Based on Counting Processes}. 
\newblock New York: Springer.
%
\bibitem[Bertozzi et~al.(2020)]{Bertozzi:et:al:20}
\newblock{Bertozzi, A. L., Franco, E., Mohler, G., Short, M. B.,  Sledge, D. (2020). The challenges of modeling and forecasting the spread of COVID-19.} 
\newblock{{\it Proceedings of the National Academy of Sciences}, 117(29), 16732-16738.}
%
\bibitem[Dean and Yang (2021)]{Dean:Yang:21}{Dean, N., and Yang, Y.} (2021).
\newblock{Discussion of “Regression Models for Understanding COVID-19 Epidemic Dynamics With Incomplete Data”,}
\newblock{\it Journal of the American Statistical Association}, {116}, 1587--1590.
%
\bibitem[Escobar(2020)]{Escobar:20}
Escobar, J. V. (2020). 
\newblock{A Hawkes process model for the propagation of COVID-19: Simple analytical results.}
\newblock{{\it Europhysics Letters}, 131(6), 68005.}
%
\bibitem[Fraser(2007)]{Fraser:07} 
Fraser, C. (2007).
\newblock{Estimating Individual and Household Reproduction Numbers in an Emerging Epidemic.}
\newblock{{\it PLoS ONE}, {2(8)}, e758.}
%
\bibitem[G\'amiz et~al.(2013)]{Gamiz:etal:13} {G\'amiz, M.~L., Janys, L., Mart{\'i}nez-Miranda, M.~D. and Nielsen, J.~P.} (2013).
\newblock{Bandwidth selection in marker dependent kernel hazard estimation,}
\newblock{\it Computational Statistics \& Data Analysis}, {68}, 155--169.
%
\bibitem[G\'amiz et~al.(2016)]{Gamiz:etal:16} {G\'amiz, M.~L., Mammen, E., Mart{\'i}nez-Miranda, M.~D. and Nielsen, J.~P.} (2016).
\newblock{Double one-sided cross-validation of local linear hazards,}
\newblock{{\it Journal of the Royal Statistical Society Series B: Statistical Methodology}, {78}, 755--779.}
%
\bibitem[G\'amiz et~al.(2022)]{Gamiz:etal:22} {G\'amiz, M.~L., Mammen, E., Mart{\'i}nez-Miranda, M.~D. and Nielsen, J.~P.} (2022).
\newblock{Missing link survival analysis with applications to available pandemic data,}
\newblock{{\it Computational Statistics \& Data Analysis}, {169}, 107405.}

\bibitem[G\'amiz et~al.(2024)]{Gamiz:etal:24}{G\'amiz, M.~L., Mammen, E., Mart{\'i}nez-Miranda, M.~D. and Nielsen, J.~P.} (2024).
\newblock{pandemics: Monitoring a Developing Pandemic with Available Data}
\newblock{R package version 0.1.0, URL  \url{https://CRAN.R-project.org/package=pandemics}}
%
\bibitem[Garetto et~al.(2021)]{Garetto:et:al21}
Garetto, M., Leonardi, E.,  Torrisi, G. L. (2021).
\newblock{ A time-modulated Hawkes process to model the spread of COVID-19 and the impact of countermeasures.}
\newblock{ {\it Annual reviews in control}, 51, 551-563.}
%

\bibitem[Garrett (2023)]{Garrett:23}{Garrett, A. (2023)}
\newblock{``12 recommendations directed at improving preparedness for the UK’s statistical ecosystem. UK Covid-19 Inquiry.''}
\newblock{Witness statement on behalf of Royal Statistical Society.}
%
\bibitem[Giudici et~al.(2023)]{Giudici:et:al:23}{Giudici, P., Pagnottoni, P., and Spelta, A.} (2023).
\newblock{  Network self-exciting point processes to measure health impacts of COVID-19.}
\newblock{\em Journal of the Royal Statistical Society Series A: Statistics in Society,} qnac006.

%
\bibitem[Goldenshluger and Koops (2019)]{Goldenshluger:Koops:19} {Goldenshluger, A. and Koops, D.} (2019).
\newblock{Nonparametric Estimation of Service Time Characteristics in Infinite-Server Queues with Nonstationary Poisson Input,}
\newblock{\it Stochastic Systems}, {9}, 1--25.
%


\bibitem[Heckman (1979)]{Heckman:79} {Heckman, J.~J.} (1979).
\newblock{Sample selection bias as a specification error,}
\newblock{\it Econometrica}, {47}, 153--161.
%
\bibitem[Jewell (2021)]{Jewell:21}{Jewell, N.~P.} (2021).
\newblock{Statistical Models for COVID-19 Incidence, Cumulative Prevalence, and $R_t$,}
\newblock{\it Journal of the American Statistical Association}, {116}, 1578--1582.
%

\bibitem[Jiang et~al. (2022)]{Jiang:et:al:22} {Jiang, F., Zhao, Z., and Shao, X.} (2022).
\newblock{ Modelling the COVID-19 infection trajectory: A piecewise linear quantile trend model.}
\newblock{\em Journal of the Royal Statistical Society Series B: Statistical Methodology}, 84(5), 1589--1607.
 %
\bibitem[Kim and Shao (2021)]{Kim:Shao:21}{ Kim, J. K., and Shao, J. }(2021).
\newblock {\em Statistical methods for handling incomplete data.}
\newblock{ CRC press.} 
%

\bibitem[Koyama et~al. (2021)]{Koyama:et:al:21}
Koyama, S., Horie, T.,  Shinomoto, S. (2021). 
\newblock{Estimating the time-varying reproduction number of COVID-19 with a state-space method.} \newblock{{\it PLoS computational biology}, 17(1), e1008679.}
\bibitem[Kresin et~al.(2022)]{Kresin:et:al:22}
Kresin, C., Schoenberg, F.,  Mohler, G. (2021). 
\newblock{Comparison of Hawkes and SEIR models for the spread of Covid-19.}
\newblock{{\it  Advances and Applications in Statistics}, 74, 83-106.}
\bibitem[Lin (2022)]{Lin:22}{Lin, X.} (2022).
\newblock{Lessons Learned from the COVID-19 Pandemic: A Statistician’s Reflection,}
\newblock{\it Statistical Science}, {37}, 278--283.
%
\bibitem[Little and Rubin (2019)] {Little:Rubin:19} {Little, R. J., and Rubin, D. B.} (2019). 
\newblock{\em Statistical analysis with missing data (Vol. 793).}
\newblock{ John Wiley \& Sons.}
%
\bibitem[Liu and Hu (2022)]{Liu:Hu:22} {Liu, Y., and Hu, F.} (2022).
\newblock{Balancing Unobserved Covariates With Covariate-Adaptive Randomized Experiments,}
\newblock{\it Journal of the American Statistical Association}, {117}(538), 875--886.
%
\bibitem[Mammen et~al.(2011)]{Mammen:etal:11} {Mammen, E., Mart{\'i}nez-Miranda, M.~D., Nielsen, J.~P. and Sperlich, S.} (2011).
\newblock{Do-validation for kernel density estimation,}
\newblock\textit{Journal of the American Statistical Association}, {106}, 651--660.
%
\bibitem[Mammen and M\"uller(2023)]{Mammen:Muller:23} {Mammen, E. and M\"uller, M.} (2023).
\newblock{Nonparametric estimation of locally stationary Hawkes processes,}
\newblock\textit{Bernoulli},  29(3), 2062--2083. 
%
\bibitem[Millimet and Parmeter (2022)] {Millimet:Parmeter:22}{Millimet, D. L., and Parmeter, C. F.} (2022).
\newblock{COVID-19 severity: A new approach to quantifying global cases and deaths.} \newblock{Journal of the Royal Statistical Society Series A: Statistics in Society}, 185(3), 1178--1215.
%
\bibitem[Mukherjee(2022)]{Mukherjee:22}{Mukherjee, B.} (2022).
\newblock{Being a Public Health Statistician During a Global Pandemic,}
\newblock{\it Statistical Science}, {37(2)}, 270--277.
%
\bibitem[Nielsen(1998)]{Nielsen:98} {Nielsen, J.~P.} (1998).
\newblock {Marker dependent kernel estimation from local linear estimation,} 
\newblock{\it Scandinavian Actuarial Journal}, {2}, 113--124.
%
\bibitem[Nash et~al.(2023)]{Nash:et:al:23}
Nash, R. K., Bhatt, S., Cori, A.,  Nouvellet, P. (2023).
\newblock{ Estimating the epidemic reproduction number from temporally aggregated incidence data: A statistical modelling approach and software tool.}
\newblock{{\it PLoS Computational Biology}, 19(8), e1011439.}
%
\bibitem[Nielsen and Linton(1995)]{Nielsen:Linton:95} {Nielsen, J.~P. and Linton, O.} (1995).
\newblock {Kernel estimation in a non-parametric marker dependent hazard model,} 
\newblock{\it The Annals of Statistics}, {23}, 1735--1748.
%
\bibitem[Park et~al. (2020)]{Park:et:al:20}{Park, S. W., Bolker, B. M., Champredon, D., Earn, D. J., Li, M., Weitz, J. S., Grenfell, B.T., and Dushoff, J.} (2020). 
\newblock {Reconciling early-outbreak estimates of the basic reproductive number and its uncertainty: framework and applications to the novel coronavirus (SARS-CoV-2) outbreak.} \newblock{\em Journal of the Royal Society Interface},  17(168), 20200144.

%
\bibitem[Pellis et~al. (2022)]{Pellis:et:al:22}{Pellis, L., Birrell, P. J., Blake, J., Overton, C. E., Scarabel, F., Stage, H. B., Danon, L., Hall, I., House, T.A., Keeling, M.J., Read, J.M., JUNIPER Consortium, and De Angelis, D.} (2022). 
\newblock{Estimation of reproduction numbers in real time: conceptual and statistical challenges.}
\newblock{\em Journal of the Royal Statistical Society Series A: Statistics in Society}, 185(Supplement\_1), S112--S130.
%

%
\bibitem[Quick et~al (2021)]{Quick:etal:21}{Quick, C., Dey, R., Lin, X.} (2021).
\newblock{ Regression Models for Understanding COVID-19 Epidemic Dynamics With Incomplete Data,}
\newblock{\it  Journal of the American Statistical Association}, {116},  1561--1577.
%
\bibitem[Richardson(2022)]{Richardson:22}{Richardson, S.} (2022). 
\newblock{Statistics in times of increasing uncertainty.}
\newblock{\em Journal of the Royal Statistical Society Series A: Statistics in Society}, 185(4), 1471--1496.

%
\bibitem[Rubin (1996)]{Rubin:96} {Rubin,D.~B.} (1996).
\newblock{Multiple imputation after 18+ years,}.
\newblock{\it Journal of the American Statistical Association}, {91}, 473--489.
%
\bibitem[Samyak and Palacios (2023)]{Samyak:Palacios:23}{Samyak, R., and Palacios, J. A.} (2023). 
\newblock{Statistical summaries of unlabelled evolutionary trees.}
\newblock{\em Biometrika}, asad025.
%
\bibitem[Stokell et~al. (2021)]{Stokell:et:al:21}{Stokell, B. G., Shah, R. D., and Tibshirani, R. J.} (2021).
\newblock{ Modelling high-dimensional categorical data using nonconvex fusion penalties.} \newblock{\em Journal of the Royal Statistical Society Series B: Statistical Methodology}, 83(3), 579--611.
%
\bibitem[Storvik et~al.(2022)]{Storvik:etal:22}{Storvik, G., Palomares, A. D. L., Engebretsen, S., Rø, G. Ø. I., Engø-Monsen, K., Kristoffersen, A. B., Freiesleben de Blasio, B., and Frigessi, A.} (2022). 
\newblock{A sequential Monte Carlo approach to estimate a time varying reproduction number in infectious disease models: the Covid-19 case.}
\newblock{arXiv preprint arXiv:2201.07590.}

%
\bibitem[Zhao et~al (2020)]{Zhao:etal:20} {Zhao, J., Kim, H.~J., Kim, H.~M.} (2020).
\newblock{New EM-type algorithms for the Heckman selection model,}
\newblock{\it Computational Statistics \& Data Analysis}, {146}, 106930.


\end{thebibliography}

%

\end{document}